\documentclass[11pt]{article} 

\usepackage[utf8]{inputenc} 

\usepackage{geometry} 
\usepackage{graphicx} 

\usepackage{tikz}
\usetikzlibrary{arrows.meta,trees, positioning}
\tikzset{step/.style={rectangle, draw, fill=blue!20, minimum height=2em, align=center}}

\usepackage{paralist} 
\usepackage{verbatim} 
\usepackage{subfig} 
\usepackage{graphicx}
\usepackage{eurosym}
\usepackage{amsmath}

\usepackage {apacite}
\usepackage{amsmath}
\usepackage{array} 
\usepackage{bibunits} 
\usepackage{booktabs} 
\usepackage{threeparttable}
\usepackage{graphicx}
\usepackage{setspace}
\usepackage[style=list,number=none]{glossary} 
\makeglossary
\usepackage{multirow}
\usepackage{setspace}
\usepackage{threeparttable}
\usepackage{tabularx}
\usepackage{enumitem}
\usepackage{float}
\usepackage{placeins}
\usepackage[font=small]{caption}
\usepackage{fancyhdr} 
\usepackage{amsfonts}
\usepackage{hyperref} 
\usepackage{sectsty}
\usepackage[titles,subfigure]{tocloft} 

\usepackage {apacite}
\usepackage{bibunits} 
\usepackage{tabularx}
\usepackage{lineno}
\title{Game Theory Analysis of Third-Party Regulation in Organic Supply Chains}
\fontsize{8pt}{12pt}\selectfont

\fontsize{12pt}{12pt}\selectfont

\author{
João Zambujal-Oliveira\\
NOVA Lincs (FCT) \\
University of Madeira\\
joliveira@staff.uma.pt\\
\and
André Silva\\
University of Madeira\\
\and
Rui Vasconcelos\\
University of Madeira\\
}

\fontsize{12pt}{12pt}\selectfont

\date{}

\begin{document}

\maketitle \thispagestyle{empty}
 \thispagestyle{empty}
\begin{abstract}
As awareness of health and environmental issues grows, the demand for organic food is rising worldwide, yet consumers still struggle to distinguish genuine organic products from conventional ones. This information asymmetry creates incentives for some producers to mislabel conventional goods as organic in order to charge higher prices, threatening market integrity and trust.

This paper develops a game-theoretic model of interactions among producers, consumers, and regulators in organic supply chains to study when fraud emerges and how it can be deterred. By analyzing extensive-form and repeated games with monitoring and penalties, we identify conditions under which honest labeling becomes a stable equilibrium and show how inspection frequency and reputation losses shape strategic behavior.

Our results highlight the critical role of a neutral third party in overcoming information asymmetries and sustaining trust in organic markets. Government regulation and independent certification, combined with credible monitoring and meaningful penalties, discourage mislabeling, and support the sustainable growth of the organic food supply chain.
%
%
%
\end{abstract}

\newpage
\begin{bibunit}[apacite]
\pagestyle{fancy} 
\setcounter{page}{1} 
\newpage
\pagenumbering{arabic}   
\singlespacing
\section{Introduction}

\begin{figure}[htp]
  \centering
  \includegraphics[width=0.8\textwidth]{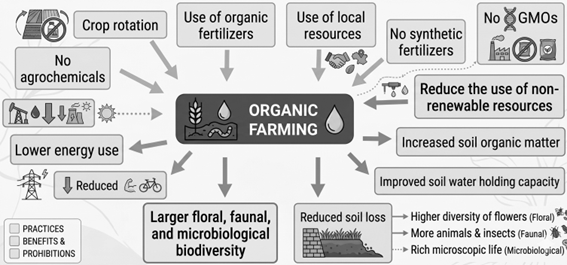}
  \caption{Key principles and benefits of organic farming. It avoids agrochemicals and synthetic fertilizers, using local resources and crop rotation. These practices reduce energy and water use, enhance biodiversity and water retention. Adapted from ~\protect\citeA{Gamage2023}.}
  \label{fig254}
\end{figure}

Organic foods have gained popularity due to their health benefits. In 2020, organic food sales in the U.S. reached \$56.4 billion, a 12.4\% increase from the previous year \cite{Organic2020}. Studies show that organic products contain higher levels of vitamin C, iron and magnesium, and and exhibit about 30\% fewer pesticide residues than conventional produce \shortcite{Brantsaeteretal2017}. Organic farming also contributes to environmental sustainability by reducing water pollution and greenhouse gas emissions \shortcite{Nejadkoorkietal2012} (Fig.~\ref{fig254}).


Environmental awareness strongly influences consumer behavior. A survey by \shortciteA{Schlegelmilchetal1996} found that 68\% of consumers consider environmental factors when buying food products. Organic farming excludes synthetic fertilizers and pesticides, relying on ecological practices \shortcite{Giampierietal2022}. However, these practices occur during production and processing, making it difficult for consumers to verify whether products are organic \cite{BournPrescott2002}. This lack of direct observability creates information asymmetry issues \cite{Mccluskey2000}.

Despite these challenges, the organic food market continues to expand. In 2022, it reached USD 208.18 billion, with a projected annual growth rate of 13\% \cite{GrandViewResearch2022}. In Europe, organic retail sales nearly doubled between 2015 and 2020, while the area under organic farming grew by 41\% \cite{EC2023}. As the market grows, economic incentives to mislabel products as organic also increase. Limited monitoring encourages opportunistic behavior by supply chain participants, who may sell conventional products as organic to gain higher prices \cite{meemkenQaim2018}.

\begin{figure}[htp]
  \centering
  \includegraphics[width=0.8\textwidth]{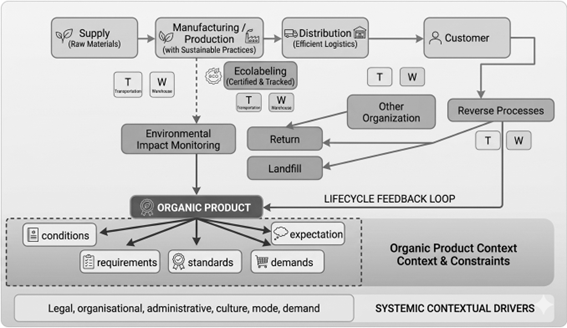}
  \caption{A closed-loop system for organic products connects supply, production, distribution, and consumption with reverse processe. Ecolabeling, standards, and regulations ensure compliance, while legal and organizational factors shape sustainability requirements. Adapted from~\protect\citeA{TundysWisniewski2020}.}
  \label{fig212}
\end{figure}

Research confirms these risks. \shortciteA{zambujal2021} studied information sharing across supply chains. \shortciteA{katsikoulietal2021} found that many supply chain agents have incentives to misreport product origins. Similarly, \shortciteA{Maetal2021} showed that such behavior destabilizes producer–consumer trust. Therefore, this paper examines how regulatory mechanisms can stabilize these interactions and foster trust within organic supply chains (Fig.~\ref{fig212}).

Recent trends, such as increasing demand for transparency \cite{ladwein2021}, rising cases of greenwashing \shortcite{mendesetal2024}, and globalized organic trade \shortcite{rathgensetal2020}, reinforce the relevance of this issue. The key challenge lies in designing monitoring systems that build trust without imposing excessive costs.

This paper explores the critical role of third-party regulation in achieving this balance. It applies game-theoretical principles to analyze interactions between producers, consumers, and regulators. Unlike earlier works that focused on supplier performance using multicriteria decision-making (MCDM) framework, this study investigates trust-building mechanisms, such as penalties, monitoring frequency, and reputation effects, to strengthen compliance. 

The analysis offers new quantitative insights into how consumers and producers interact in organic markets. Building on the work of \shortciteA{krishnanetal2021} and \shortciteA{Maetal2021}, it explores whether effective regulation should depend on consumer monitoring or independent third-party supervision. It also examines how reputation influences the balance of trust among market participants \shortcite{lauetal2020}.

From these, three hypotheses emerge: 
1) higher monitoring frequency and penalties improve regulatory compliance; 
2) consumer-driven reputation loss increases the likelihood of organic production; and 
3) third-party involvement enhances trust among all supply chain agents.

By analyzing simple one-shot games, this paper identifies incentives that help build trust across the supply chain \shortcite{haringetal2009,OrganicTargets4EU2026}. The research questions focus on what penalty levels can discourage deceptive behavior and how penalties and monitoring can be combined for effective regulation. They also examine how consumer-imposed reputation loss influences producers’ willingness to comply and how third-party regulation reinforces trust among supply chain agents.


\section{Existing Research on Organic Supply Chain}
\subsection{Regulatory Mechanisms and Market Dynamics}
Information asymmetry between producers and consumers is central to organic food markets. \shortciteA{Wanetal2012} and \shortciteA{Zhaoetal2025} classified products as search or credence goods depending on how easily consumers can verify quality.
Building on this framework, \citeA{Mccluskey2000} analyzed interactions between producers and consumers using sequential games. He showed that opportunistic producers may exploit consumers when organic attributes are hard to verify. 

In his model, monitoring intensity depends on economic parameters. Higher monitoring increases the profitability of organic products and widens the cost gap with conventional production, but its effectiveness declines with higher discount rates. \shortciteA{Amatoetal2015} confirmed that environmental claims are difficult to validate without monitoring, reinforcing the credence nature of organic goods.

\begin{table}[htbp]
  \centering
  \small
  \scalebox{0.65}{
    \begin{tabular}{|p{11.775em}|p{11.775em}|p{11.775em}|p{11.775em}|p{11.775em}|}
    \toprule
    Author(s) & Goal & Methods & Conclusions & Strengths/Limitations \\
    \midrule
    \citeA{Mccluskey2000} & Examine incentives under asymmetric information & Sequential games; Policy analysis & Monitoring rises with organic profitability and production cost gaps; declines with higher discount rates & Clear model; limited payoff notation \\
    \midrule
    \shortciteA{lauetal2020} & Identify factors ensuring truthful organic labeling & Strategy games; Field data analysis & Regular and random government monitoring needed; laws and penalties reinforce compliance & Uses real data; complex structure \\
    \midrule
    \citeA{ZhangGeorgescu2022} & Assess conditions for sustainable organic supply chains & Evolutionary games; Simulation models & Consumer income, awareness, and subsidies drive stability and growth & Diverse methods; lacks empirical validation \\
    \bottomrule
    \end{tabular}%
  }
  \caption{Game-theoretical approaches to organic supply chains under asymmetric information}
  \label{tab315}%
\end{table}%

\shortciteA{lauetal2020} extended this analysis to the entire supply chain. Using competitive game-theoretic models and field data, they identified incentives among producers, retailers, and consumers.
They found that random and regular inspections are essential to prevent opportunistic behavior. However, they also showed that an equilibrium rarely emerges between suppliers and retailers because verification costs are high.
These findings underline the role of government monitoring, legislation, and penalties in maintaining trust. Table \ref{tab315} compares major studies linking regulation and market performance in organic supply chains.

\begin{figure}[H]
  \centering
   \includegraphics[width=0.8\textwidth]{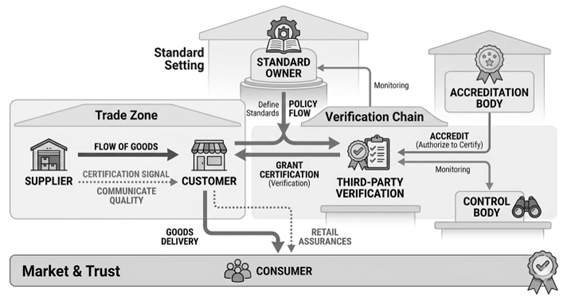}
  \caption{Structure of a certification and accreditation system. Standards guide suppliers, while certification and accreditation ensure compliance. Consumers benefit from trusted quality through monitored interactions. Adapted from \protect\citeA{Jahnetal2005}.}\label{fig325}
\end{figure}

The inspection framework proposed by \shortciteA{Jahnetal2005} (Fig. \ref{fig325}) illustrates how multiple agents interact during certification. Failures may occur at any stage, justifying the need for periodic inspections across all agents \shortcite{Albersmeieretal2009,Houssiere2024}.

Government agencies also influence market equilibria through subsidies and sanctions. \citeA{ZhangGeorgescu2022} modeled the government as an active player enforcing standards and shaping consumer utility.
Their results show that subsidies and environmental awareness programs can shift purchasing behavior toward sustainable products, stabilizing the organic market.
%
%
\subsection{Game Theory Applications in Organic Supply Chain Regulation}
Game theory provides a structured way to analyze strategic interactions in organic food supply chains.
It helps evaluate how regulatory interventions affect market behavior and trust among producers, regulators, and consumers.
\shortciteA{taghikhahetal2021} demonstrated that these models can predict how monitoring and certification policies influence verification rates, adoption of organic practices, and certification times. 

Game-theoretic models allow regulators to test policy alternatives and anticipate behavioral responses.
According to \shortciteA{Richetal2011} and \citeA{GhoshShah2012}, this approach clarifies how each agent’s payoff depends on others’ actions, revealing where trust and cooperation may fail.

\shortciteA{lauetal2020} conducted a case study modeling interactions between suppliers and supermarkets. 
A mixed-strategy equilibrium occurs when the retailer monitors 39.5\% of the time and the supplier produces organic goods with a 36.4\% probability. This partial compliance implies that consumers may unknowingly purchase non-organic products. The result illustrates a market failure arising from weak monitoring incentives.
%

\begin{figure}[htbp]
  \centering
  \includegraphics[width=0.6\textwidth]{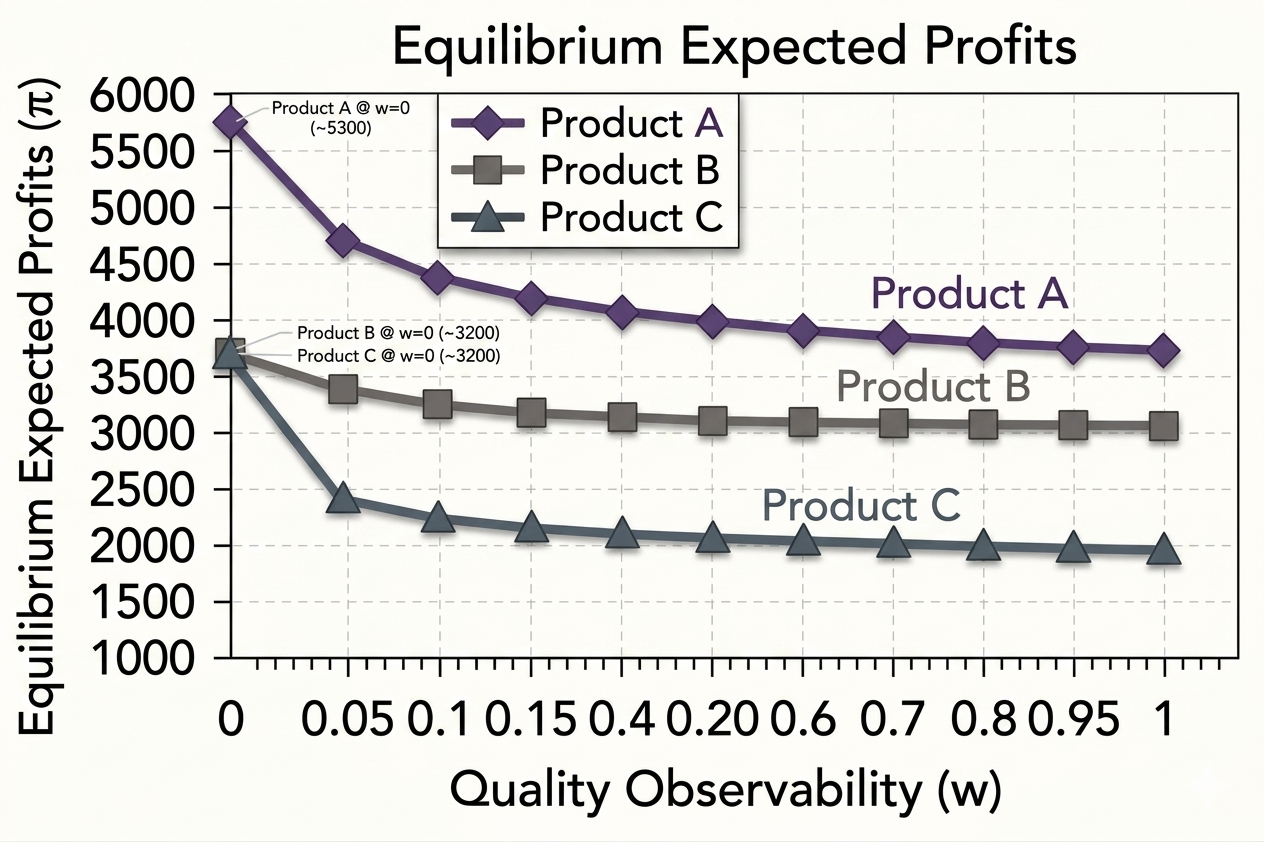}
  \caption{Quality consumer observation vs. firm's expected profits. Greater observability reduces information asymmetry, while transparency and monitoring promote fairness and market stability \protect{\cite{CarriquiryBabcock2007}}.}
  \label{fig584}
\end{figure}

As shown in Fig. \ref{fig584}, higher consumer recognition of quality reduces distributors’ expected profits \cite{CarriquiryBabcock2007,Zhangetal2025}. \citeA{NoelkeCaswell2000} observed that some retailers exploit brand reputation to signal quality without proper verification. When monitoring costs are high, retailers may compromise authenticity. Thus, government oversight becomes essential to prevent fraud and sustain consumer trust \cite{ManziniAccorsi2013}. When consumers face high monitoring costs, they tend not to verify producers’ claims \cite{AndersenPhilipsen1998}. Under these conditions, producers benefit from falsely labeling non-organic products as organic. This behavior constitutes the only stable Nash equilibrium \cite{ManningKowalska2021}.

As long as information remains asymmetric and monitoring costly, fraudulent practices persist \shortcite{Thogersenetal2017}. \shortciteA{Yuetal2021} examined how producers decide whether to adopt organic production when retailers share demand information. They found that conversion to organic practices increases when production costs decrease and consumer preference strengthens. However, cooperation between producers and retailers remains unstable, highlighting the need for regulatory intervention.%
%
Market maturity also limits organic sector growth. \shortciteA{Melovic2020} argued that distribution bottlenecks restrict expansion, while \citeA{Lockie2006} noted that high price differentials maintain organic products as a niche market. Government regulation plays a decisive role in overcoming these barriers.
Subsidies for early adopters and consumer awareness campaigns can enhance trust \shortcite{ZhangGeorgescu2022}. However, \citeA{Giannakas2002} found that mislabeling certification weakens trust. Nevertheless, there is evidence that sustainable packaging positively contributes to circular food supply chains \cite{zambujal2024}. Using a regulatory game model, \citeA{Jia2011} demonstrated that strong government oversight improves compliance. Policies that raise penalties for illegal production, reduce regulatory costs, and increase official wages can maximize social welfare \shortcite{Golanetal2001}.

Overall, game theory helps identify equilibria where monitoring, incentives, and information flow determine market integrity.
It reveals that regulatory design—particularly government monitoring and penalty structures—is essential to sustain trust and efficiency in organic supply chains.

\section{Regulation Model for the Organic Supply Chain}
\subsection{Extensive Form Games}
A sequential game can be represented in extensive form as a decision tree. It consists of a set of players $(N = \{1, 2, \dots, n\})$ who traverse edges $(E)$ connecting nodes$(E)$ $(E \subseteq X \times X)$. Thus, edges represent moves between nodes $(X)$. At certain nodes, players can make choices $(D \subseteq X)$, while at others—disjoint from the previous ones—the game ends and payoffs are assigned $(Z \subseteq X, \quad D \cap Z = \emptyset)$. 
Each decision node has a player function that assigns a player to that node $([P: D \rightarrow N])$ where $(P(x) = i)$ means player $(i)$ makes the decision at node $(x)$.   

Additionally, each decision node has a set of available actions, denoted by $(A(x))$.  At each terminal node, there is a payoff function that assigns a payoff vector to that node $u: Z \rightarrow \mathbb{R}^n$ where $u(z) = (u_1(z), u_2(z), \dots, u_n(z))$ and $(u_i(z))$ represents the payoff to player $(i)$ at terminal node $(z)$.  

\subsection{Backward Induction Algorithm}
The backward induction algorithm is used to solve finite perfect-information games by working backward from the terminal nodes. After identifying the terminal nodes and their payoffs, the optimal action at each decision node is determined to maximize the player's payoff. If $(h)$ is a terminal node, assign its payoff as $(v(h) = u_i(h))$. Otherwise, if $(h)$ is a decision node, compute:
\begin{equation}
    v(h) = \max_{a \in A(h)} u_i(f(h,a)),
\end{equation}
where $A(h)$ is the set of available actions at $(h)$, and $(f(h,a))$ is the resulting node when action $(a)$ is taken. By iteratively replacing each decision node with its best possible value, this process continues until reaching the root node, which will contain the optimal strategy profile.
\subsection{Regulation Models for the Organic Supply Chain}
This methodological framework for the organic food supply chain aims to analyze interactions between producers and consumers dynamically. To study multiple outcomes and strategies, a structured approach is needed. Extensive decision trees are well-suited for this purpose.

Frequent interactions between producers and consumers require a dynamic analysis \shortciteA{Rosenthal1981}. The relationship can be modeled as a two-party, extensive, alternating game. \citeA{FudenbergTirole1989} agree that decision alternatives, information, and pay-offs form a decision tree. This approach captures strategies for consuming both organic and conventional foods.

The baseline scenario assumes the consumer decides whether to monitor the origin of organic food. The second step includes the retailer in the monitoring process. The third step adds a third party to ensure monitoring, even probabilistically. In this context, the game involves a producer (\emph{P}) and a consumer (\emph{C}). The producer may attempt to defraud the consumer, creating strategic misalignment. Before producing organic foods and making claims, the producer is uncertain about potential monitoring. Likewise, the consumer is unsure about the truth of organic claims without monitoring.

\subsubsection{Consumer Monitoring in Organic Food Markets}
\begin{figure}[htbp]
  \centering
  \includegraphics[width=0.7\textwidth]{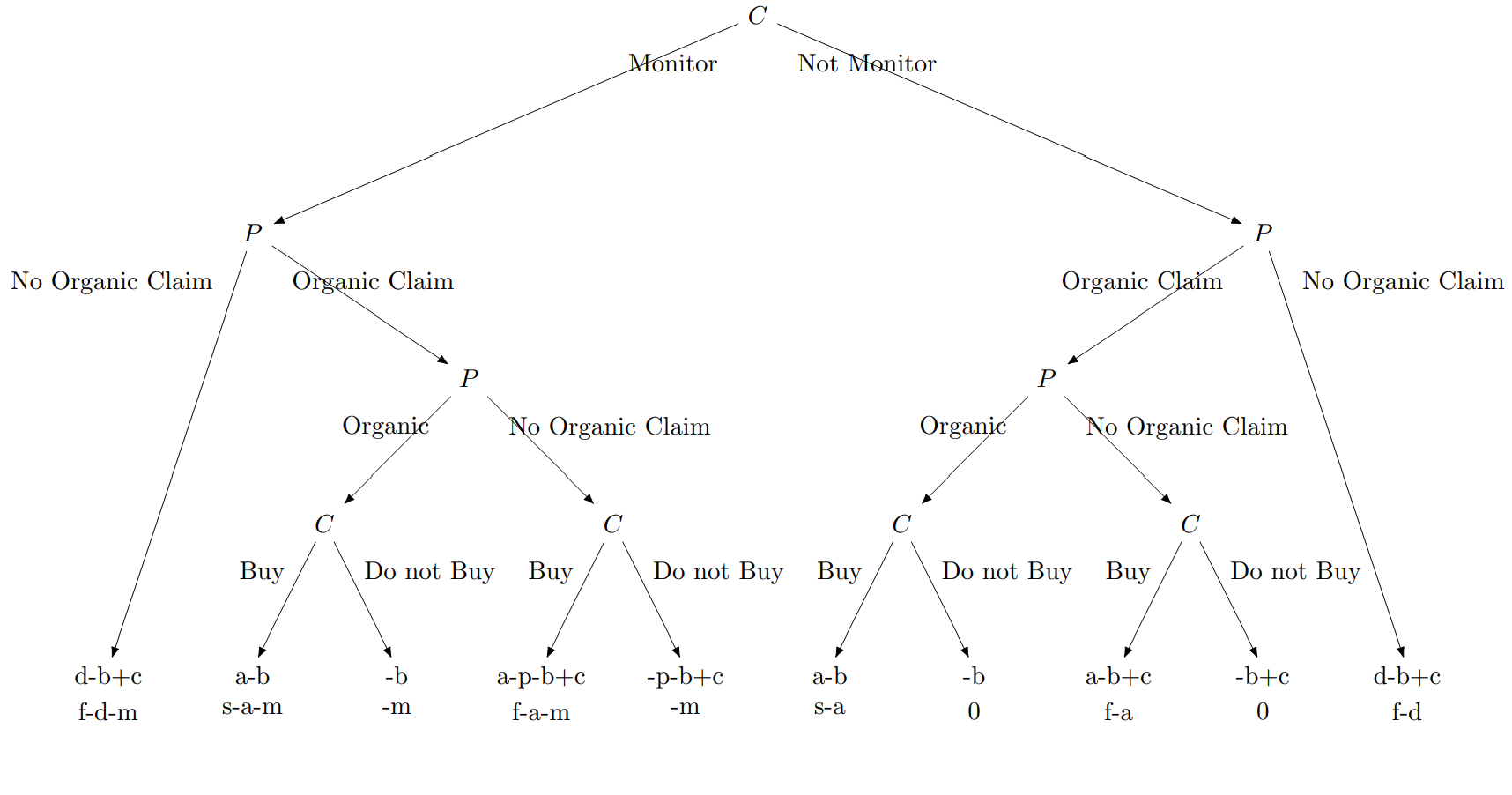}
  \caption{Producer vs. Consumer with Consumer Monitoring.}\label{fig324}
\end{figure}

At the initial stage, the model assumes a limited governmental role, which may include intervention in the producer-consumer relationship by penalizing fraudulent producers (where the penalty may involve reputational damage). The analysis disregards strategies that do not involve organic claims, as these lack relevant game dynamics. The sequential and extensive representation of the game is given in Fig.~\ref{fig324}.

\begin{center}
\scriptsize
\begin{tabular}{ll}
$P$: producer; & $C$: consumer; \\
$p$: penalty for fraud; & $a$: price of organic food; \\
$d$: price of non-organic food; & $s$: utility from organic consumption; \\
$f$: utility from conventional consumption; & $m$: cost of monitoring; \\
$b$: cost of organic production; & $c$: cost difference between organic and conventional farming.
\end{tabular}
\normalsize
\end{center}

The model incorporates several assumptions regarding the payoff structure. First, the price of organically produced foods exceeds or equals that of conventional ones:
\begin{gather}
a \geq d
\end{gather}
Second, the utility derived from organic food consumption, denoted \( s \), is equal to that of conventional food \( f \) when the consumer cannot distinguish between them: \( s = f \). Otherwise, if the consumer can tell the difference, then \( s > f \). From Fig.~\ref{fig324}, and as discussed by \shortciteA{Chenetal2007}, the strategy "Don't Buy" is strictly dominated by "Buy" for the consumer:
\begin{enumerate}[label=\roman*., itemsep=0pt, parsep=0pt,topsep=0pt]
    \item Monitor and Organic Claim:\\
    $s - a - m > -m$ $\Rightarrow$ Buy $>$ Don't Buy (Organic) \\
    $f - d > 0$  $\Rightarrow$ \text{Buy $>$ Don't Buy (Not Organic)} 
      \item Not Monitor and Organic Claim:\\
      $s - a > 0$  $\Rightarrow$ \text{Buy $>$ Don't Buy (Organic)} \\
     $f - d > 0$  $\Rightarrow$ \text{Buy $>$ Don't Buy (Not Organic)}
\end{enumerate}
Therefore, the strategy “Don’t Buy” is eliminated from the consumer's strategy set. Aware of potential unethical business practices \shortcite{Maetal2021} or aiming to improve trust in the organic food supply chain \cite{Segerson1999}, the government may intervene. It can appoint third-party agents to perform random monitoring. The regulatory penalty \( p \), as proposed by \shortciteA{Maetal2021}, is a fine for producers caught falsely marketing non-organic products as organic. This penalty rises proportionally with the gains from deception. To prevent producers from making false “Organic Claims” while selling non-organic products, the honest payoff must be greater than the deceptive one.

\begin{gather}
a - b > a - p - (b - c) \Rightarrow p > c \label{eq526}
\end{gather}

After presenting the game model, the next step is to determine the penalty \(p\) and monitoring cost \(m\). These must ensure that the producer offers genuine organic products and that the consumer chooses to monitor and buy. To define a \textit{Subgame Perfect Nash Equilibrium (SPNE)}, we identify conditions where each player's strategy is optimal in every subgame. This yields three equilibrium conditions:

\begin{enumerate}[label=\roman*., itemsep=0pt, parsep=0pt,topsep=0pt]
    \item Consumer buys when the utility from consuming organic food exceeds its price: \( s > a \);
    \item Producer tells the truth when the penalty outweighs the gain from cheating: \( p > c \);
    \item Consumer monitors when the benefit of distinguishing organic from conventional outweighs the monitoring cost: \( m < s - f \).
\end{enumerate}
From these conditions, we derive the equilibrium thresholds:
\begin{enumerate}[label=\roman*., itemsep=0pt, parsep=0pt,topsep=0pt]
    \item Minimum penalty to deter cheating: \( p^* > c \);
    \item Maximum monitoring cost to incentivize oversight: \( m^* < s - f \).
\end{enumerate}

These thresholds define the parameter space where truthful labeling by the producer and active monitoring by the consumer constitute a game equilibrium.

\subsection{Consumer Monitoring in Organic Food Markets With Reputational Loss}

The model now incorporates monitoring by an intermediary in the supply chain, the retailer. Even accounting for monitoring costs (passed to consumers), the difference is the introduction of a reputational loss \( t \) when the retailer is exposed as selling falsely labeled products. According to \citeA{Giannakas2002}, a market-based penalty such as \( t \) arises when consumers detect deception and reject the product. This leads to a loss of market share, functioning as an endogenous punishment mechanism.
\begin{figure}[htbp]
  \centering
  \includegraphics[width=0.7\textwidth]{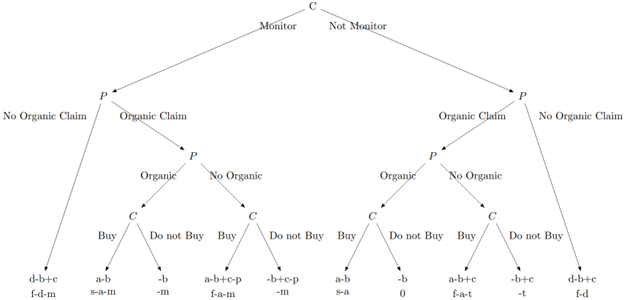}
  \caption{Producer vs. Consumer payoff matrix, with consumer-financed monitoring}
  \label{fig689B}

  \begin{center}
  \scriptsize
  \begin{tabular}{ll}
  $P$: producer; & $C$: consumer; \\
  $p$: penalty; & $a$: price for organic food; \\
  $d$: price for non-organic food; & $s$: utility obtained from organic consumption; \\
  $f$: utility from conventional consumption; & $m$: cost of monitoring; \\
  $r$: probability of successful enforcement; & $b$: cost of organic production;\\
  $c$: cost of conventional production.
  \end{tabular}
  \normalsize
  \end{center}
\end{figure}

Several insights can be drawn from the analysis of Fig. \ref{fig689B}. To identify the Nash equilibria, we evaluate the conditions under which each player chooses their best response, depending on whether monitoring occurs. When the monitoring occurs, the producer has an incentive to produce honestly if the payoff of selling genuine organic products exceeds that from cheating and risking a penalty \( p \).
  \begin{gather}
    a - b > -p - m \Rightarrow p > a - b + m
    \tag{1}
  \end{gather}
  Thus, the penalty must exceed the net gain from cheating, adjusted for the monitoring cost. In other words, the consumer's incentive to buy derives from the difference in utility between buying organic and not buying, given that monitoring costs are sunk.
  \begin{gather}
    s - a - m > -m \Rightarrow s > a
  \end{gather}
In the absence of monitoring, a producer has an incentive to cheat because a dishonest producer will earn:
  \begin{gather}
    a - c > a - b \Rightarrow c < b
  \end{gather}
  Since organic production is more costly (\( b > c \)), cheating is preferred. If the consumer believes the product is falsely labeled, their utility depends on the perceived reputational loss \( t \) when deceived:
  \begin{gather}
    f - a - t < 0 \Rightarrow t > f - a
  \end{gather}
  This implies that reputational loss must exceed the consumer's net benefit from potentially mislabeled products. Without monitoring, the game has no stable Nash equilibrium. The producer will cheat, and the consumer, anticipating this, will choose not to buy. When monitoring is introduced and the penalty \(p\) is high enough to deter cheating, a Nash equilibrium emerges. 
  
  In this equilibrium, the producer sells genuine organic products, and the consumer purchases them. This result aligns with \shortciteA{Zhaoetal2020}, who show that reliable monitoring can restore trust in the organic food supply chain. By contrast, \shortciteA{lauetal2020} found no equilibrium without monitoring, highlighting the importance of third-party support.

\subsection{Third-Party Monitoring in Organic Food Markets}

\begin{figure}[htbp]
  \centering
  \includegraphics[width=0.7\textwidth]{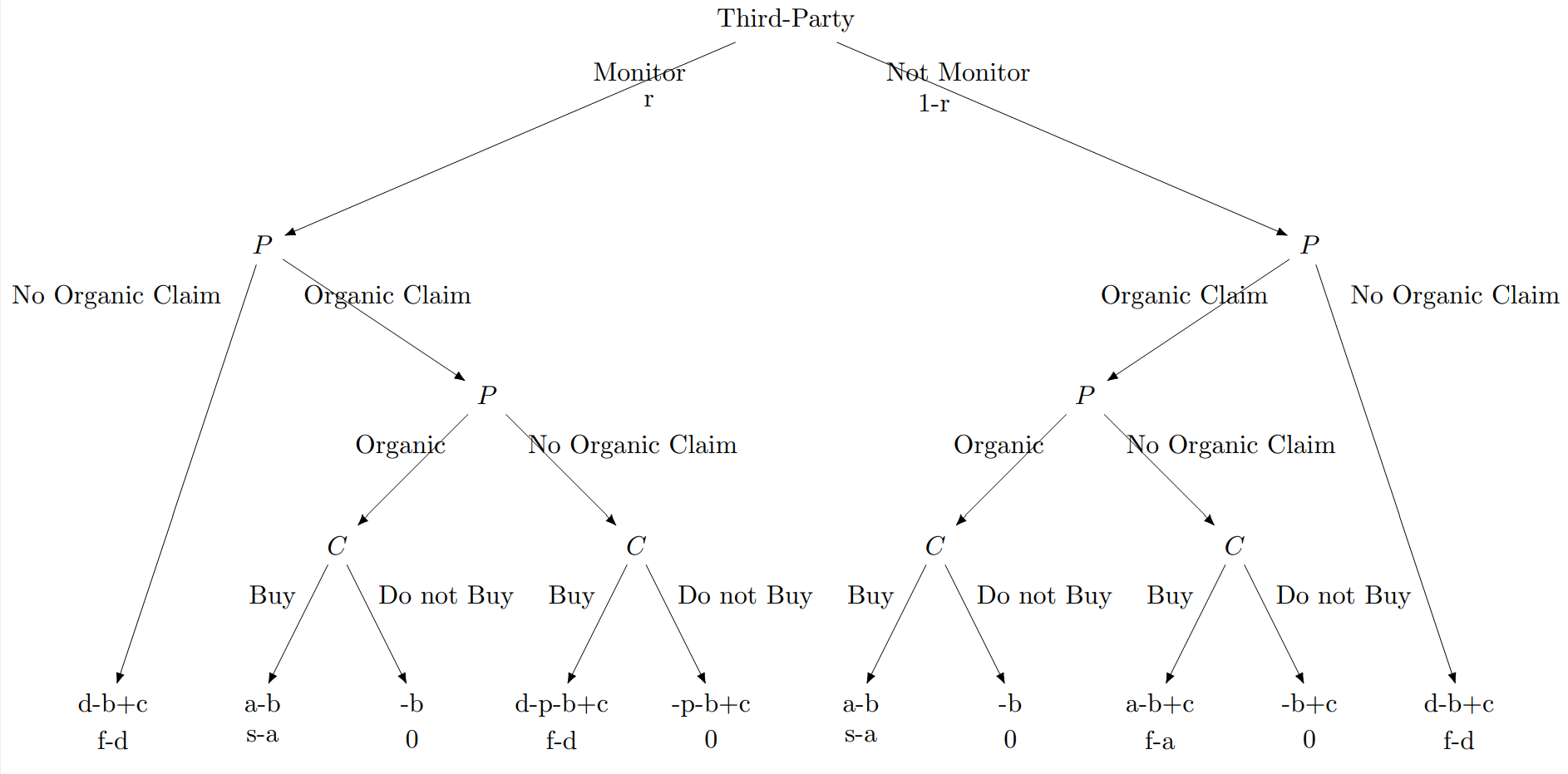}
  \caption{Producer vs. consumer with third-party monitoring}
  \label{fig587}
  \begin{center}
  \scriptsize
  \begin{tabular}{ll}
  $P$: producer; & $C$: consumer; \\
  $p$: penalty; & $a$: price for organic food; \\
  $d$: price for non-organic food; & $s$: utility obtained from organic consumption; \\
  $f$: utility obtained from conventional consumption; & $m$: cost of monitoring; \\
  $r$: probability of being accurately regulated; & $b - c$: cost difference between organic and conventional farming.
 \end{tabular}
  \normalsize
  \end{center}
\end{figure}

The third development of the framework entails shifting the burden of monitoring from the consumer to a third party. This change eliminates the previous constraint, wherein consumers bore the cost of monitoring, but it introduces an element of uncertainty into the process. Producers are now subject to random monitoring, with a probability \( r \) of effective regulation (Fig.~\ref{fig587}).

The aim of this model is to deter the producer from falsely claiming ``organic'' when the product is not. An honest producer earns a payoff of:
\begin{equation}
    a - b
\end{equation}

A dishonest producer, on the other hand, expects the following payoff:
\begin{equation}
    r(d - p - b + c) + (1 - r)(a - b + c)
\end{equation}

The incentive compatibility condition---requiring that honesty yields at least as much payoff as dishonesty---is:
\begin{equation}
    a - b \geq r(d - p - b + c) + (1 - r)(a - b + c)
\end{equation}

To ensure this outcome, we seek a combination of \( r \) (monitoring probability) and \( p \) (penalty) that satisfies three conditions:
\begin{enumerate}[label=\roman*., itemsep=0pt, parsep=0pt,topsep=0pt]
    \item The consumer chooses to buy when the product is truly organic,
    \item The producer prefers to be honest when making an organic claim,
    \item The producer avoids falsely claiming ``organic'' when the product is not.
\end{enumerate}

To encourage consumers to buy organic and deter dishonest produc ers, we must ensure consumers value organic products more than their price:
\begin{equation}
    s > a
\end{equation}

Moreover, to satisfy the producer’s incentive compatibility, the following condition on \( p \) and \( r \) must hold:
\begin{equation}
    p \geq \frac{d + (1 - r)(a - b + c) - (a - b)}{r}
\end{equation}

This implies that the higher the monitoring probability \( r \), the lower the required penalty \( p \) to deter dishonesty. Conversely, if monitoring is infrequent (i.e., \( r \) is small), a higher penalty is needed to prevent cheating.

It becomes increasingly apparent that the consumer's payoffs improve with third-party regulation. Payoffs for consuming regulated products are now at least positive, and potentially even higher than those from consuming non-regulated products. One could reasonably assume an increase in utility from the certainty about a product’s nature (e.g., a regulator-issued label). As confirmed by \shortciteA{Aschemannetal2007}, the strategy "Buy" for consuming regulated products strictly dominates the strategy of not buying altogether.

However, consumers will not choose to purchase food revealed as non-organic if the utility of consuming confirmed non-organic foods, \( f - d \), is lower than the utility of consuming non-revealed products sold as organic, \( s - a \). This is because when the truthfulness of the claim is not revealed:
\begin{equation}
    f = d \Rightarrow (f - d) \geq (s - a)
\end{equation}

Thus, the consumer's optimal strategies involve purchasing goods that are either certain or at least perceived as organic. The producer achieves the best payoffs by selling organic goods when monitored and by selling non-organic foods when not monitored. These two scenarios represent competitive equilibria in the game.

Based on these statements, a formulation can be developed for the probability \( r \) of achieving accurate regulation:

\begin{equation}
    E = \left\{ (a, b, c, r) \in \Omega: r > \frac{2a - 2b + c}{a - b + c} \right\}
\end{equation}

\begin{center}
\scriptsize
\( \Omega \): set for all possible values of the real-valued random variables concerning the producer.
\normalsize
\end{center}

To mitigate or disincentivize illicit practices by the producer \shortcite{Vanetal2020}, the probability of being monitored should be higher than the payoff generated by deceiving the consumer and should be a proportion of the sum of the payoffs from the two producer strategies (selling organic and non-organic products). Therefore, as the payoff obtained from selling non-organic products as organic increases, the probability of being monitored must also increase. These findings align with the solution to repeated normal-form games proposed by \citeA{Mccluskey2000} 

\section{Case Analysis}

We now illustrate the model with a numerical example and determine equilibrium values for the key parameters. Consider the following values:

\begin{table}[htp]
\centering
\scriptsize
\caption{Baseline parameter values and empirical calibration}
\label{tab578}
\begin{tabularx}{\textwidth}{
  p{0.05\textwidth} 
  p{0.05\textwidth} 
  p{0.06\textwidth} 
  p{0.45\textwidth} 
  p{0.25\textwidth}
}
\toprule
\textbf{(1)} & \textbf{(2)} & \textbf{(3)} & \textbf{(4)} & \textbf{(5)} \\
\midrule

\(a\) & 12 & 10--20 
& Calibrated using retail scanner data, where organic products are typically priced 10–100
& \citeA{dimitrietal2002, USDAERS2025}; \\

\(d\) & 8  & 5--10  
& Based on retail benchmarks and price variation across food categories, ensuring a realistic organic price premium.
& \shortciteA{USDAERS2016,EC2019,Mulleretal2019} \\

\(b\) & 7  & 5--12  
& Derived from farm-level cost studies showing higher organic production costs; the baseline uses a mid-range estimate.
& \citeA{FAO2017,Zanolietal2012}\\

\(c\) & 4  & 2--6   
& Represents conventional or mislabeling costs, which are lower due to the absence of certification and regulatory requirements.
& \shortciteA{Lusk2011,EC2019,TundysWisniewski2020} \\

\(s\) & 14 & 10--25 
& Calibrated from willingness-to-pay estimates for organic attributes. Interpreted as reduced-form utility consistent with experimental and survey-based evidence. 
& \shortciteA{LoureiroHine2001,Johnstonetal2017}\\

\(f\) & 8  & 6--10  
& Captures consumer utility under imperfect information. Empirical studies show lower valuation when organic quality is not verifiable. 
& \citeA{Giletal2000, Bontempsetal2013,Leeetal2024}\\

\bottomrule
\end{tabularx}

\vspace{0.2cm}
\footnotesize
\begin{quote}
\textit{Notes:} (1) Parameter; (2) Baseline value; (3) Empirical range; (4) Calibration methodology; (5) Empirical references.
Parameter values are calibrated using a combination of retail price data, farm‑level cost studies, and willingness‑to‑pay estimates from the organic food literature. 
\end{quote}
\end{table}

The parameterization follows a calibration approach combining retail price data, farm-level cost evidence, and experimental willingness-to-pay estimates. 
This ensures consistency between market outcomes and behavioral foundations, enhancing the external validity of the model.

An honest producer earns a payoff of:
\begin{gather}
a - b = 12 - 7 = 5
\end{gather}

A dishonest producer earns an expected payoff of:
\begin{gather}
r(d - p - b + c) + (1 - r)(a - b + c) = r(5 - p) + (1 - r)(9)
\end{gather}

The incentive compatibility condition becomes:
\begin{align}
5 &\geq 5r - rp + 9(1 - r) \\
4r + rp &\geq 4
\end{align}

Solving for the minimum penalty \( p \):
\begin{gather}
p \geq \frac{4(1 - r)}{r}
\end{gather}

\noindent
We now compute the minimum required penalty \( p \) for selected values of the monitoring probability \( r \):

\begin{table}[htp]
\centering
\begin{tabular}{ccc}
\toprule
$r$ & Expression for $p$ & Minimum $p$ \\
\midrule
0.2 & $\frac{4(1 - 0.2)}{0.2}$ & 16 \\
0.4 & $\frac{4(1 - 0.4)}{0.4}$ & 6 \\
0.6 & $\frac{4(1 - 0.6)}{0.6}$ & $\approx 2.67$ \\
0.8 & $\frac{4(1 - 0.8)}{0.8}$ & 1 \\
1.0 & $\frac{4(1 - 1.0)}{1.0}$ & 0 \\
\bottomrule
\end{tabular}
\caption{Minimum penalty $p$ required to deter dishonest behavior}
\end{table}
These results illustrate that a higher probability of monitoring reduces the need for a large penalty. When monitoring is perfect (\( r = 1 \)), no penalty is required. We also verify the consumer incentive condition \((14 > 12)\) confirming that consumers are willing to buy organic products when they trust the claim. To evaluate the probabilistic condition for achieving equilibrium:
\begin{gather}
r > \frac{2(12) - 2(7) + 4}{12 - 7 + 4} = \frac{14}{9} \approx 1.56
\end{gather}
This result indicates that no feasible monitoring probability \( r \in [0, 1] \) can satisfy the condition, and thus other mechanisms such as repeated-game strategies or higher penalties must be used to ensure compliance. 

\section{Sensitivity Analysis}
To assess the robustness of the numerical findings and the stability of regulatory equilibria in the organic supply chain model, we perform a one-way sensitivity analysis on the six baseline parameters. 
We vary each parameter individually by -40\%, -20\%, +20\%, and +40\% and recompute the minimum \( p \) at representative monitoring probabilities \( r = 0.4 \) and \( r = 0.8 \), as well as the critical \( r \) threshold. 
%
\begin{table}[!htbp]
\centering
\scriptsize
\caption{Impact of Parameter Variations on Minimum Penalty \(p\) and Critical Threshold \(r^*\)}
\label{tab748}
\begin{tabular}{l c c c c}
\toprule
\textbf{Parameter} & \textbf{Variation} & \textbf{Value} & \textbf{Min. Penalty \(p\)} & \textbf{Critical \(r^*\)} \\
\midrule

\multirow{4}{*}{\(a\) (organic price)}
& -40\% & 7.2  & 4.8 (r=0.4), 0.8 (r=0.8) & 0.89 \\
& -20\% & 9.6  & 7.2, 1.2 & 1.22 \\
& +20\% & 14.4 & 8.4, 1.4 & 1.89 \\
& +40\% & 16.8 & — & 2.22 \\

\midrule
\multirow{4}{*}{\(d\) (non-organic price)}
& -40\% & 4.8  & 5.4, 0.9 & 1.44 \\
& -20\% & 6.4  & 5.7, 0.95 & 1.50 \\
& +20\% & 9.6  & 6.3, 1.05 & 1.61 \\
& +40\% & 11.2 & — & 1.67 \\

\midrule
\multirow{4}{*}{\(b\) (organic cost)}
& -40\% & 4.2  & 3.6, 0.6 & 1.11 \\
& -20\% & 5.6  & 4.8, 0.8 & 1.33 \\
& +20\% & 8.4  & 7.2, 1.2 & 1.78 \\
& +40\% & 9.8  & — & 2.00 \\

\midrule
\multirow{4}{*}{\(c\) (non-organic cost)}
& -40\% & 2.4  & 4.8, 0.8 & 1.33 \\
& -20\% & 3.2  & 5.4, 0.9 & 1.44 \\
& +20\% & 4.8  & 6.6, 1.1 & 1.67 \\
& +40\% & 5.6  & — & 1.78 \\

\midrule
\multirow{4}{*}{\(s\) (utility organic)}
& -40\% & 8.4  & \multicolumn{2}{c}{No direct impact (consumer side)} \\
& -20\% & 11.2 & \multicolumn{2}{c}{No direct impact (consumer side)} \\
& +20\% & 16.8 & \multicolumn{2}{c}{No direct impact (consumer side)} \\
& +40\% & 19.6 & \multicolumn{2}{c}{No direct impact (consumer side)} \\

\midrule
\multirow{4}{*}{\(f\) (utility non-organic)}
& -40\% & 4.8  & \multicolumn{2}{c}{Minor effect} \\
& -20\% & 6.4  & \multicolumn{2}{c}{Minor effect} \\
& +20\% & 9.6  & \multicolumn{2}{c}{Minor effect} \\
& +40\% & 11.2 & \multicolumn{2}{c}{Minor effect} \\

\bottomrule
\end{tabular}
\end{table}

\begin{figure}[htp]
  \centering
  \includegraphics[width=\textwidth]{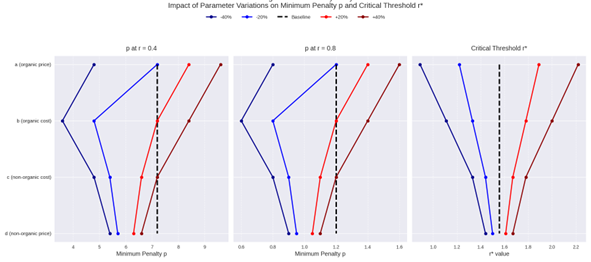}
  \caption{Impact of parameter variations.}
  \label{fig820}
\end{figure}

Table~\ref{tab748} and Figure~\ref{fig820} allow the following interpretation of the results. Increasing Price of organic food\( a \) (higher premium) widens the incentive to cheat, requiring higher penalties or near-perfect monitoring. A 40\% increase raises the critical \( r \)-threshold above 2.0, making equilibrium even harder to achieve without strong third-party intervention. This aligns with McCluskey (2000), who highlights how larger price gaps exacerbate asymmetric information problems in credence goods markets.

Higher conventional prices of non-organic food (\(d\)) reduce the relative gains from fraudulent behavior and therefore lower the penalties required to sustain compliance. However, this effect is modest compared to the influence of production costs. By contrast, the cost of organic production (\(b\)) plays a relevant role in shaping cheating incentives. 

Higher organic costs, due to certification requirements and ecological production methods, increase the attractiveness of fraud. In particular, a 40\% increase in \(b\) drives the critical monitoring threshold \(r^*\) beyond feasible levels. This reinforces the importance of cost-reducing policies and is consistent with evolutionary game-theoretic analyses. 

The cost of non-organic production (\(c\)) is also influential. A reduction in \(c\), which increases the cost advantage of cheating, contributes strongly to destabilizing compliance. Although a 40\% decrease in \(c\) slightly lowers the critical monitoring threshold, it dramatically increases the required penalties when monitoring is weak. This is consistent with findings that illegal production yields higher net benefits under lax regulation.

Changes in the utility derived from organic food consumption (\(s\)) primarily affect consumer-side incentives rather than producer penalty thresholds. Higher consumer valuation strengthens demand for verified organic products and makes purchasing a dominant strategy, while lower values of \(s\) increase the risk of market collapse even in the presence of regulatory enforcement. 

Finally, the utility from non-organic food (\(f\)) influences consumers’ willingness to accept non-verified products or engage in monitoring. While lower values of \(f\) widen the perceived utility gap and support monitoring, their direct impact on producer-side equilibrium conditions remains limited.
In general terms, the model is most sensitive to the cost differential (\( b - c \)) and the organic price premium (\( a - d \)). Small increases in fraud profitability (via higher \( b \), lower \( c \), or higher \( a \)) push the required monitoring probability beyond 1.0 or demand unrealistically high penalties. This confirms the paper’s emphasis on third-party regulation: when market parameters favor cheating, consumer-only monitoring fails, and third-party oversight with appropriate \( r \) and \( p \) becomes essential \shortcite{lauetal2020}.

Under realistic ranges, third-party monitoring probabilities above 60–70\% combined with moderate penalties (or reputational losses) can restore compliance. However, the baseline scenario’s unfeasible critical \( r > 1.56 \) highlights the necessity of complementary mechanisms such as repeated interactions, subsidies to reduce \( b \), or certification systems that effectively raise consumer utility \( s \).

\section{Discussion}

The sensitivity analysis demonstrated that the regulatory recommendations (higher \( r \), calibrated \( p \), and support for reputation effects) are robust in direction but highly parameter-dependent in magnitude. 

The game-theoretical model provides insights into how regulation and agents interact to deter fraudulent labeling in organic food supply chains. By quantifying penalty thresholds that discourage misrepresentation, it offers policymakers a tool to adjust interventions according to market conditions. Therefore, policymakers should prioritize reducing the organic cost premium through targeted subsidies and invest in monitoring infrastructure to achieve \( r \geq 0.7 \) whenever cost asymmetries are large.

The model shows that stricter regulatory environments substantially reduce dishonest behavior. This outcome supports \citeA{Mccluskey2000} and \citeA{StarbirdAmanor-Boadu2007}, who found that consistent penalties and third-party monitoring enhance consumer trust. In contrast, lenient settings with low penalties or rare inspections lead to weaker deterrence and riskier equilibria. These findings stress the need to calibrate both penalties and monitoring frequency to match the level of market risk.

Monitoring frequency strongly influences producer incentives. Consistent with \shortciteA{Ehmkeetal2019} and \citeA{Sahay2003}, our results show that third-party monitoring raises consumer confidence but cannot fully eliminate fraud when expected profits remain high. For instance, if inspection probability is only 10\%, producers may still mislabel products when the potential profit increases by 50\%. This supports \citeA{BondarevaPinker2019}, who observed that inspection rates below 10\% significantly increase non-compliance.

Reputational effects further strengthen compliance. The model estimates that mislabeling can lead to a 30–50\% loss in consumer trust, consistent with \shortciteA{kroetzetal2020}, who reported that reputation-based penalties reduce fraud risk by up to 60\%. Thus, combining monitoring with reputational consequences encourages honest behavior, even without excessive regulation.

\begin{figure}[htp]
  \centering
  \includegraphics[width=0.8\textwidth]{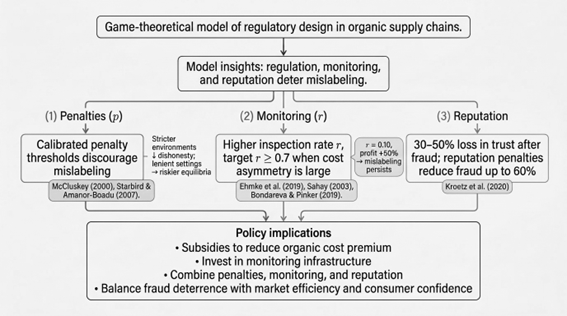}
  \caption{Model of regulatory design in organic supply chains.}
  \label{fig863}
\end{figure}

Overall, the model clarifies how penalties, monitoring, and reputation interact to sustain credibility and trust in organic supply chains. This framework supports the design of balanced policies that deter fraud while maintaining market efficiency and consumer confidence (Figure~\ref{fig863}).

\section{Policy Implications}

In fact, our case analysis evidences that organic supply chains require some form of inspection and monitoring. Considering the cost asymmetry
between suppliers and consumers, the competitive equilibria defined by our game model can contribute to balancing the negotiation power within organic food supply chains and markets. Therefore, our game theory model's new cost-benefit analyses of proposed policies can demonstrate their potential impact and value (Table \ref{tab767}).

\begin{table}[htbp]
\scriptsize
\centering
\caption{Policy Levers and Effects on Market Behavior}
\begin{tabular}{|l|p{10cm}|}
\hline
\textbf{Policy Lever} & \textbf{Effect on Market Behavior} \\
\hline
Increase $r$ & Reduces required penalty, increases deterrence of dishonesty. \\
\hline
Increase $p$ & Compensates for lower $r$, but may be politically or economically constrained. \\
\hline
Labeling/Certification & Increases consumer utility, further incentivizing purchase of organic goods. \\
\hline
Cost of Monitoring ($m$) & Should be weighed against benefits of increased market honesty. \\
\hline
\end{tabular}
\label{tab767}%
\end{table}
The model underscores the critical role of third-party monitoring. External institutions that bear monitoring costs can enhance equilibrium outcomes. This supports \citeA{Mccluskey2000}, who argue that both government and private institutions can effectively perform this role. However, as \shortciteA{lauetal2020} note, private regulators may become complacent, emphasizing the need for independent oversight.

\begin{figure}[htp]
\centering
\includegraphics[width=0.7\textwidth]{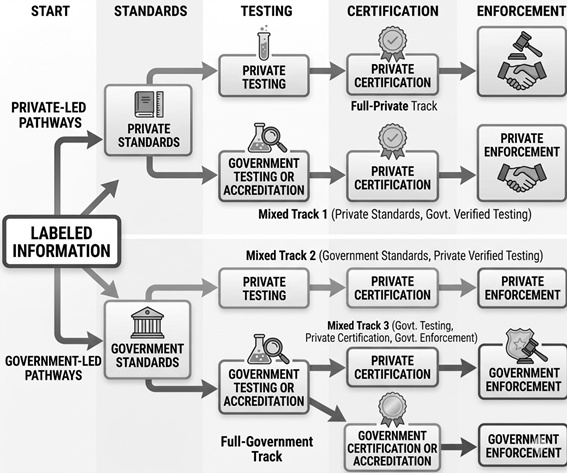}
\caption{Different labeling configurations. Adapted from~\protect\citeA{Golanetal2001}.}
\label{fig659}
\end{figure}

Empirical evidence supports \shortciteA{Hatanakaetal2005} and \shortciteA{baueretal2022}, who show that third-party certifications reshape competitive equilibria and influence agents’ behavior in the supply chain. Exploiting variation in certification thresholds on platforms such as eBay, \shortciteA{baueretal2022} demonstrate that stricter quality standards increase seller effort and shift the quality distribution. Nevertheless, other authors such as \citeA{SanogoMasters2002} and \citeA{Fagan2003} have questioned their contribution to market efficiency. Our model does not distinguish between the roles of government and private institutions in improving trust. In such cases, multiple labeling schemes, as proposed by \shortciteA{Golanetal2001} (Fig. \ref{fig659}), may provide differentiated policy solutions.

Related work on green-bond certification by \citeA{xing2025} further suggests that third-party rating policies can influence post-issuance innovation and disclosure, reinforcing the importance of design-aware certification regimes in current environmental-finance frameworks. Our case analysis could not resolve the questions raised by \shortciteA{Gunninghametal1999} about when third parties act as surrogate regulators and how governments can facilitate their effectiveness.

Targeted policies focusing on labeling and certification visibility can further strengthen consumer confidence \cite{janssenHamm2012}. Partnerships to share certification best practices also improve collaboration among supply chain agents \cite{gerardi2023}. Certifications can increase organic product prices by about 10\% \cite{lohr1998}, raising revenue for certified producers. However, certification costs may reach up to 5\% of revenues for small agents \shortciteA{yuetal2022}, creating barriers to entry and limiting competition.
 
Finally, the relevance of third-party regulation is reinforced by technologies such as blockchain-based traceability \shortcite{katsikoulietal2021}. Combined with periodic audits and labeling policies, these mechanisms can enhance trust and market efficiency in organic food systems.%
%
\section{Conclusions}
This study examined the role of third-party regulation in ensuring the credibility of organic food supply chains. It identified the policy measures most effective in sustaining trust and preventing fraudulent labeling. Using a game-theoretical approach, the study quantified the penalty range needed to discourage producers from misrepresenting conventional products as organic.

The strategic and extensive-form models revealed that regulatory oversight is essential. Because consumers face high monitoring costs, they often rely on suppliers’ organic claims. The model shows that as monitoring costs increase, consumer trust declines, emphasizing the need for inspection systems and third-party certification. These findings are consistent with \citeA{BondarevaPinker2019}, who observed that in the absence of inspections, producers have strong incentives to deviate from honest behavior.

The main contribution lies in identifying regulatory levers that enhance trustworthiness in organic supply chains. The model defines penalty levels and monitoring probabilities that support stable equilibria between producers and consumers. It also outlines governance mechanisms, such as participatory certification systems, that can strengthen stakeholder engagement.

The results further suggest that optimizing monitoring frequency improves regulatory accuracy and credibility. These insights can inform policy decisions aimed at balancing inspection costs with the benefits of consumer confidence and fair competition. Some limitations provide directions for future work. The model assumes fixed monitoring probabilities and rational expectations. 

Future studies could introduce uncertainty in producers’ beliefs about being inspected or in consumers’ confidence in organic labels \shortcite{Jahnetal2005, janssenHamm2012}. Additional research could also address consumer behavior factors such as brand loyalty and willingness to pay for verified organic products \shortcite{baquero2022, akaichietal2012}.
\putbib[manuscript.2510.12420V2.bib]
\end{bibunit}


\begin{thebibliography}{}

\bibitem [\protect \citeauthoryear {%
Akaichi%
, Nayga%
\BCBL {}\ \BBA {} Gil%
}{%
Akaichi%
\ \protect \BOthers {.}}{%
{\protect \APACyear {2012}}%
}]{%
akaichietal2012}
\APACinsertmetastar {%
akaichietal2012}%
\begin{APACrefauthors}%
Akaichi, F.%
, Nayga, R\BPBI M., Jr%
\BCBL {}\ \BBA {} Gil, J\BPBI M.%
\end{APACrefauthors}%
\unskip\
\newblock
\APACrefYearMonthDay{2012}{}{}.
\newblock
{\BBOQ}\APACrefatitle {Assessing consumers' willingness to pay for different
  units of organic milk: evidence from multiunit auctions} {Assessing
  consumers' willingness to pay for different units of organic milk: evidence
  from multiunit auctions}.{\BBCQ}
\newblock
\APACjournalVolNumPages{Canadian Journal of Agricultural
  Economics}{60}{4}{469--494}.
\newblock
\begin{APACrefDOI} \doi{10.1111/j.1744-7976.2012.01254.x} \end{APACrefDOI}
\PrintBackRefs{\CurrentBib}

\bibitem [\protect \citeauthoryear {%
Albersmeier%
, Schulze%
, Jahn%
\BCBL {}\ \BBA {} Spiller%
}{%
Albersmeier%
\ \protect \BOthers {.}}{%
{\protect \APACyear {2009}}%
}]{%
Albersmeieretal2009}
\APACinsertmetastar {%
Albersmeieretal2009}%
\begin{APACrefauthors}%
Albersmeier, F.%
, Schulze, H.%
, Jahn, G.%
\BCBL {}\ \BBA {} Spiller, A.%
\end{APACrefauthors}%
\unskip\
\newblock
\APACrefYearMonthDay{2009}{}{}.
\newblock
{\BBOQ}\APACrefatitle {The reliability of third-party certification in the food
  chain: From checklists to risk-oriented auditing} {The reliability of
  third-party certification in the food chain: From checklists to risk-oriented
  auditing}.{\BBCQ}
\newblock
\APACjournalVolNumPages{Food Control}{20}{10}{927--935}.
\newblock
\begin{APACrefDOI} \doi{10.1016/j.foodcont.2009.01.009} \end{APACrefDOI}
\PrintBackRefs{\CurrentBib}

\bibitem [\protect \citeauthoryear {%
Amato%
, Zillante%
\BCBL {}\ \BBA {} Amato%
}{%
Amato%
\ \protect \BOthers {.}}{%
{\protect \APACyear {2015}}%
}]{%
Amatoetal2015}
\APACinsertmetastar {%
Amatoetal2015}%
\begin{APACrefauthors}%
Amato, L\BPBI H.%
, Zillante, A.%
\BCBL {}\ \BBA {} Amato, C\BPBI H.%
\end{APACrefauthors}%
\unskip\
\newblock
\APACrefYearMonthDay{2015}{}{}.
\newblock
{\BBOQ}\APACrefatitle {Corporate environmental claims: a game theory model with
  empirical results} {Corporate environmental claims: a game theory model with
  empirical results}.{\BBCQ}
\newblock
\APACjournalVolNumPages{Social Responsibility Journal}{11}{1}{36--55}.
\newblock
\begin{APACrefDOI} \doi{10.1108/SRJ-05-2013-0058} \end{APACrefDOI}
\PrintBackRefs{\CurrentBib}

\bibitem [\protect \citeauthoryear {%
Andersen%
\ \BBA {} Philipsen%
}{%
Andersen%
\ \BBA {} Philipsen%
}{%
{\protect \APACyear {1998}}%
}]{%
AndersenPhilipsen1998}
\APACinsertmetastar {%
AndersenPhilipsen1998}%
\begin{APACrefauthors}%
Andersen, E.%
\BCBT {}\ \BBA {} Philipsen, K.%
\end{APACrefauthors}%
\unskip\
\newblock
\APACrefYearMonthDay{1998}{}{}.
\newblock
{\BBOQ}\APACrefatitle {The evolution of credence goods in customer markets:
  exchanging 'pigs in pokes'} {The evolution of credence goods in customer
  markets: exchanging 'pigs in pokes'}.{\BBCQ}
\newblock
\APACjournalVolNumPages{DRUID Winter Seminar}{}{}{}.
\PrintBackRefs{\CurrentBib}

\bibitem [\protect \citeauthoryear {%
Aschemann%
, Hamm%
, Naspetti%
\BCBL {}\ \BBA {} Zanoli%
}{%
Aschemann%
\ \protect \BOthers {.}}{%
{\protect \APACyear {2007}}%
}]{%
Aschemannetal2007}
\APACinsertmetastar {%
Aschemannetal2007}%
\begin{APACrefauthors}%
Aschemann, J.%
, Hamm, U.%
, Naspetti, S.%
\BCBL {}\ \BBA {} Zanoli, R.%
\end{APACrefauthors}%
\unskip\
\newblock
\APACrefYearMonthDay{2007}{}{}.
\newblock
{\BBOQ}\APACrefatitle {The organic market} {The organic market}.{\BBCQ}
\newblock
\BIn{} \APACrefbtitle {Organic Farming: An International History} {Organic
  farming: An international history}\ (\BPGS\ 123--151).
\newblock
\APACaddressPublisher{}{CABI, Wallingford, UK}.
\newblock
\begin{APACrefDOI} \doi{10.1079/9780851998336.0123} \end{APACrefDOI}
\PrintBackRefs{\CurrentBib}

\bibitem [\protect \citeauthoryear {%
Baquero%
}{%
Baquero%
}{%
{\protect \APACyear {2022}}%
}]{%
baquero2022}
\APACinsertmetastar {%
baquero2022}%
\begin{APACrefauthors}%
Baquero, A.%
\end{APACrefauthors}%
\unskip\
\newblock
\APACrefYearMonthDay{2022}{}{}.
\newblock
{\BBOQ}\APACrefatitle {{Net promoter score (NPS) and customer satisfaction:
  relationship and efficient management}} {{Net promoter score (NPS) and
  customer satisfaction: relationship and efficient management}}.{\BBCQ}
\newblock
\APACjournalVolNumPages{Sustainability}{14}{4}{2011}.
\newblock
\begin{APACrefDOI} \doi{10.3390/su14042011} \end{APACrefDOI}
\PrintBackRefs{\CurrentBib}

\bibitem [\protect \citeauthoryear {%
Bauer%
, Parra-Moyano%
, Schmedders%
\BCBL {}\ \BBA {} Schwabe%
}{%
Bauer%
\ \protect \BOthers {.}}{%
{\protect \APACyear {2022}}%
}]{%
baueretal2022}
\APACinsertmetastar {%
baueretal2022}%
\begin{APACrefauthors}%
Bauer, I.%
, Parra-Moyano, J.%
, Schmedders, K.%
\BCBL {}\ \BBA {} Schwabe, G.%
\end{APACrefauthors}%
\unskip\
\newblock
\APACrefYearMonthDay{2022}{}{}.
\newblock
{\BBOQ}\APACrefatitle {Multi-party certification on blockchain and its impact
  in the market for lemons} {Multi-party certification on blockchain and its
  impact in the market for lemons}.{\BBCQ}
\newblock
\APACjournalVolNumPages{Journal of Management Information
  Systems}{39}{2}{395--425}.
\newblock
\begin{APACrefDOI} \doi{10.1016/j.irfa.2025.104536} \end{APACrefDOI}
\PrintBackRefs{\CurrentBib}

\bibitem [\protect \citeauthoryear {%
Bondareva%
\ \BBA {} Pinker%
}{%
Bondareva%
\ \BBA {} Pinker%
}{%
{\protect \APACyear {2019}}%
}]{%
BondarevaPinker2019}
\APACinsertmetastar {%
BondarevaPinker2019}%
\begin{APACrefauthors}%
Bondareva, M.%
\BCBT {}\ \BBA {} Pinker, E.%
\end{APACrefauthors}%
\unskip\
\newblock
\APACrefYearMonthDay{2019}{}{}.
\newblock
{\BBOQ}\APACrefatitle {Dynamic relational contracts for quality enforcement in
  supply chains} {Dynamic relational contracts for quality enforcement in
  supply chains}.{\BBCQ}
\newblock
\APACjournalVolNumPages{Management Science}{65}{3}{1305--1321}.
\newblock
\begin{APACrefDOI} \doi{10.1287/mnsc.2017.2990} \end{APACrefDOI}
\PrintBackRefs{\CurrentBib}

\bibitem [\protect \citeauthoryear {%
Bontemps%
, Busson%
\BCBL {}\ \BBA {} Carlier%
}{%
Bontemps%
\ \protect \BOthers {.}}{%
{\protect \APACyear {2013}}%
}]{%
Bontempsetal2013}
\APACinsertmetastar {%
Bontempsetal2013}%
\begin{APACrefauthors}%
Bontemps, C.%
, Busson, H.%
\BCBL {}\ \BBA {} Carlier, A.%
\end{APACrefauthors}%
\unskip\
\newblock
\APACrefYearMonthDay{2013}{}{}.
\newblock
{\BBOQ}\APACrefatitle {Information, quality and certification: The case of
  French cheese labels} {Information, quality and certification: The case of
  french cheese labels}.{\BBCQ}
\newblock
\APACjournalVolNumPages{European Review of Agricultural
  Economics}{40}{1}{1--35}.
\newblock
\begin{APACrefDOI} \doi{10.1093/erae/jbs025} \end{APACrefDOI}
\PrintBackRefs{\CurrentBib}

\bibitem [\protect \citeauthoryear {%
Bourn%
\ \BBA {} Prescott%
}{%
Bourn%
\ \BBA {} Prescott%
}{%
{\protect \APACyear {2002}}%
}]{%
BournPrescott2002}
\APACinsertmetastar {%
BournPrescott2002}%
\begin{APACrefauthors}%
Bourn, D.%
\BCBT {}\ \BBA {} Prescott, J.%
\end{APACrefauthors}%
\unskip\
\newblock
\APACrefYearMonthDay{2002}{}{}.
\newblock
{\BBOQ}\APACrefatitle {A comparison of the nutritional value, sensory
  qualities, and food safety of organically and conventionally produced foods}
  {A comparison of the nutritional value, sensory qualities, and food safety of
  organically and conventionally produced foods}.{\BBCQ}
\newblock
\APACjournalVolNumPages{Critical Reviews in Food Science and
  Nutrition}{42}{1}{1--34}.
\newblock
\begin{APACrefDOI} \doi{10.1080/10408690290825439} \end{APACrefDOI}
\PrintBackRefs{\CurrentBib}

\bibitem [\protect \citeauthoryear {%
Brants{\ae}ter%
, Ydersbond%
, Hoppin%
, Haugen%
\BCBL {}\ \BBA {} Meltzer%
}{%
Brants{\ae}ter%
\ \protect \BOthers {.}}{%
{\protect \APACyear {2017}}%
}]{%
Brantsaeteretal2017}
\APACinsertmetastar {%
Brantsaeteretal2017}%
\begin{APACrefauthors}%
Brants{\ae}ter, A\BPBI L.%
, Ydersbond, T\BPBI A.%
, Hoppin, J\BPBI A.%
, Haugen, M.%
\BCBL {}\ \BBA {} Meltzer, H\BPBI M.%
\end{APACrefauthors}%
\unskip\
\newblock
\APACrefYearMonthDay{2017}{}{}.
\newblock
{\BBOQ}\APACrefatitle {Organic food in the diet: exposure and health
  implications} {Organic food in the diet: exposure and health
  implications}.{\BBCQ}
\newblock
\APACjournalVolNumPages{Annual Review of Public Health}{38}{}{295--313}.
\newblock
\begin{APACrefDOI} \doi{10.1146/annurev-publhealth-031816-044437}
  \end{APACrefDOI}
\PrintBackRefs{\CurrentBib}

\bibitem [\protect \citeauthoryear {%
Carriquiry%
\ \BBA {} Babcock%
}{%
Carriquiry%
\ \BBA {} Babcock%
}{%
{\protect \APACyear {2007}}%
}]{%
CarriquiryBabcock2007}
\APACinsertmetastar {%
CarriquiryBabcock2007}%
\begin{APACrefauthors}%
Carriquiry, M.%
\BCBT {}\ \BBA {} Babcock, B\BPBI A.%
\end{APACrefauthors}%
\unskip\
\newblock
\APACrefYearMonthDay{2007}{}{}.
\newblock
{\BBOQ}\APACrefatitle {Reputations, market structure, and the choice of quality
  assurance systems in the food industry} {Reputations, market structure, and
  the choice of quality assurance systems in the food industry}.{\BBCQ}
\newblock
\APACjournalVolNumPages{American Journal of Agricultural
  Economics}{89}{1}{12--23}.
\newblock
\begin{APACrefDOI} \doi{10.1111/j.1467-8276.2007.00959.x} \end{APACrefDOI}
\PrintBackRefs{\CurrentBib}

\bibitem [\protect \citeauthoryear {%
Chen%
, Van~Long%
\BCBL {}\ \BBA {} Luo%
}{%
Chen%
\ \protect \BOthers {.}}{%
{\protect \APACyear {2007}}%
}]{%
Chenetal2007}
\APACinsertmetastar {%
Chenetal2007}%
\begin{APACrefauthors}%
Chen, Y\BHBI C.%
, Van~Long, N.%
\BCBL {}\ \BBA {} Luo, X.%
\end{APACrefauthors}%
\unskip\
\newblock
\APACrefYearMonthDay{2007}{}{}.
\newblock
{\BBOQ}\APACrefatitle {Iterated Strict Dominance in General Games} {Iterated
  strict dominance in general games}.{\BBCQ}
\newblock
\APACjournalVolNumPages{Games and Economic Behavior}{61}{2}{299--315}.
\newblock
\begin{APACrefDOI} \doi{10.1016/j.geb.2006.10.009} \end{APACrefDOI}
\PrintBackRefs{\CurrentBib}

\bibitem [\protect \citeauthoryear {%
Dimitri%
, Greene%
\BCBL {}\ \protect \BOthers {.}}{%
Dimitri%
\ \protect \BOthers {.}}{%
{\protect \APACyear {2002}}%
}]{%
dimitrietal2002}
\APACinsertmetastar {%
dimitrietal2002}%
\begin{APACrefauthors}%
Dimitri, C.%
, Greene, C.%
\BCBL {}\ \BOthersPeriod {.}\end{APACrefauthors}%
\unskip\
\newblock
\APACrefYearMonthDay{2002}{}{}.
\newblock
{\BBOQ}\APACrefatitle {Recent growth patterns in the US organic foods market}
  {Recent growth patterns in the us organic foods market}.{\BBCQ}
\newblock
\APACjournalVolNumPages{Organic Agriculture in the US}{}{}{129--190}.
\PrintBackRefs{\CurrentBib}

\bibitem [\protect \citeauthoryear {%
{EC}%
}{%
{EC}%
}{%
{\protect \APACyear {2019}}%
}]{%
EC2019}
\APACinsertmetastar {%
EC2019}%
\begin{APACrefauthors}%
{EC}.%
\end{APACrefauthors}%
\unskip\
\newblock
\APACrefYearMonthDay{2019}{}{}.
\newblock
\APACrefbtitle {Report on Food Fraud and Mislabelling in EU Food Markets}
  {Report on food fraud and mislabelling in eu food markets}\ (Report\ \BNUM\
  SANTE/2018/123456).
\newblock
\APACrefnote{{European Commission}}
\PrintBackRefs{\CurrentBib}

\bibitem [\protect \citeauthoryear {%
{ECDGARD}%
}{%
{ECDGARD}%
}{%
{\protect \APACyear {2023}}%
}]{%
EC2023}
\APACinsertmetastar {%
EC2023}%
\begin{APACrefauthors}%
{ECDGARD}.%
\end{APACrefauthors}%
\unskip\
\newblock
\APACrefYearMonthDay{2023}{January}{}.
\newblock
\APACrefbtitle {{Organic farming in the EU – A decade of organic growth}}
  {{Organic farming in the EU – A decade of organic growth}}\ \APACbVolEdTR
  {}{Report}.
\newblock
\APACaddressInstitution{}{European Commission}.
\PrintBackRefs{\CurrentBib}

\bibitem [\protect \citeauthoryear {%
Ehmke%
, Bonanno%
, Boys%
\BCBL {}\ \BBA {} Smith%
}{%
Ehmke%
\ \protect \BOthers {.}}{%
{\protect \APACyear {2019}}%
}]{%
Ehmkeetal2019}
\APACinsertmetastar {%
Ehmkeetal2019}%
\begin{APACrefauthors}%
Ehmke, M\BPBI D.%
, Bonanno, A.%
, Boys, K.%
\BCBL {}\ \BBA {} Smith, T\BPBI G.%
\end{APACrefauthors}%
\unskip\
\newblock
\APACrefYearMonthDay{2019}{}{}.
\newblock
{\BBOQ}\APACrefatitle {Food fraud: Economic insights into the dark side of
  incentives} {Food fraud: Economic insights into the dark side of
  incentives}.{\BBCQ}
\newblock
\APACjournalVolNumPages{Australian Journal of Agricultural and Resource
  Economics}{63}{4}{685--700}.
\newblock
\begin{APACrefDOI} \doi{10.1111/1467-8489.12346} \end{APACrefDOI}
\PrintBackRefs{\CurrentBib}

\bibitem [\protect \citeauthoryear {%
Fagan%
}{%
Fagan%
}{%
{\protect \APACyear {2003}}%
}]{%
Fagan2003}
\APACinsertmetastar {%
Fagan2003}%
\begin{APACrefauthors}%
Fagan, J.%
\end{APACrefauthors}%
\unskip\
\newblock
\APACrefYearMonthDay{2003}{}{}.
\newblock
{\BBOQ}\APACrefatitle {Cert ID, a successful example of an independent,
  third-party, private certification system} {Cert id, a successful example of
  an independent, third-party, private certification system}.{\BBCQ}
\newblock
\BIn{} \APACrefbtitle {Symposium ``Product Differentiation and Market
  Segmentation in Grains and Oilseeds: Implications for Industry in
  Transition''} {Symposium ``product differentiation and market segmentation in
  grains and oilseeds: Implications for industry in transition''}\ (\BPGS\
  27--28).
\newblock
\APACaddressPublisher{Washington, DC}{Economic Research Service, USDA and The
  Farm Foundation}.
\newblock
\begin{APACrefURL}
  \url{https://www.farmfoundation.org/wp-content/uploads/attachments/235-Fagan.pdf}
  \end{APACrefURL}
\newblock
\APACrefnote{Conference paper; no DOI assigned}
\PrintBackRefs{\CurrentBib}

\bibitem [\protect \citeauthoryear {%
FAO%
}{%
FAO%
}{%
{\protect \APACyear {2017}}%
}]{%
FAO2017}
\APACinsertmetastar {%
FAO2017}%
\begin{APACrefauthors}%
FAO.%
\end{APACrefauthors}%
\unskip\
\newblock
\APACrefYearMonthDay{2017}{}{}.
\newblock
\APACrefbtitle {Comparative Analysis of Organic and Non-Organic Farming
  Systems} {Comparative analysis of organic and non-organic farming systems}\
  \APACbVolEdTR{}{\BTR{}}.
\newblock
\APACaddressInstitutionEqAuth{Rome, Italy}{FAO}.
\newblock
\APACrefnote{{Food and Agriculture Organization of the United Nations}}
\PrintBackRefs{\CurrentBib}

\bibitem [\protect \citeauthoryear {%
Fudenberg%
\ \BBA {} Tirole%
}{%
Fudenberg%
\ \BBA {} Tirole%
}{%
{\protect \APACyear {1989}}%
}]{%
FudenbergTirole1989}
\APACinsertmetastar {%
FudenbergTirole1989}%
\begin{APACrefauthors}%
Fudenberg, D.%
\BCBT {}\ \BBA {} Tirole, J.%
\end{APACrefauthors}%
\unskip\
\newblock
\APACrefYearMonthDay{1989}{}{}.
\newblock
{\BBOQ}\APACrefatitle {Noncooperative game theory for industrial organization:
  an introduction and overview} {Noncooperative game theory for industrial
  organization: an introduction and overview}.{\BBCQ}
\newblock
\BIn{} R.~Schmalensee\ \BBA {} R.~Willig\ (\BEDS), \APACrefbtitle {Handbook of
  Industrial Organization} {Handbook of industrial organization}\ (\BVOL~1,
  \BPGS\ 259--327).
\newblock
\APACaddressPublisher{}{Elsevier}.
\PrintBackRefs{\CurrentBib}

\bibitem [\protect \citeauthoryear {%
Gamage%
\ \protect \BOthers {.}}{%
Gamage%
\ \protect \BOthers {.}}{%
{\protect \APACyear {2023}}%
}]{%
Gamage2023}
\APACinsertmetastar {%
Gamage2023}%
\begin{APACrefauthors}%
Gamage, A.%
, Gangahagedara, R.%
, Gamage, J.%
, Jayasinghe, N.%
, Kodikara, N.%
, Suraweera, P.%
\BCBL {}\ \BBA {} Merah, O.%
\end{APACrefauthors}%
\unskip\
\newblock
\APACrefYearMonthDay{2023}{}{}.
\newblock
{\BBOQ}\APACrefatitle {Role of organic farming for achieving sustainability in
  agriculture} {Role of organic farming for achieving sustainability in
  agriculture}.{\BBCQ}
\newblock
\APACjournalVolNumPages{Farming System}{1}{1}{100005}.
\newblock
\begin{APACrefDOI} \doi{10.1016/j.farsys.2023.100005} \end{APACrefDOI}
\PrintBackRefs{\CurrentBib}

\bibitem [\protect \citeauthoryear {%
Gerardi%
}{%
Gerardi%
}{%
{\protect \APACyear {2023}}%
}]{%
gerardi2023}
\APACinsertmetastar {%
gerardi2023}%
\begin{APACrefauthors}%
Gerardi, A.%
\end{APACrefauthors}%
\unskip\
\newblock
\APACrefYearMonthDay{2023}{}{}.
\newblock
{\BBOQ}\APACrefatitle {{Global Food Safety Initiative (GFSI): underpinning the
  safety of the global food chain, facilitating regulatory compliance, trade,
  and consumer trust}} {{Global Food Safety Initiative (GFSI): underpinning the
  safety of the global food chain, facilitating regulatory compliance, trade,
  and consumer trust}}.{\BBCQ}
\newblock
\BIn{} \APACrefbtitle {Present Knowledge in Food Safety} {Present knowledge in
  food safety}\ (\BPGS\ 1089--1098).
\newblock
\APACaddressPublisher{}{Elsevier}.
\newblock
\begin{APACrefDOI} \doi{10.1016/B978-0-12-819470-6.00068-9} \end{APACrefDOI}
\PrintBackRefs{\CurrentBib}

\bibitem [\protect \citeauthoryear {%
Ghosh%
\ \BBA {} Shah%
}{%
Ghosh%
\ \BBA {} Shah%
}{%
{\protect \APACyear {2012}}%
}]{%
GhoshShah2012}
\APACinsertmetastar {%
GhoshShah2012}%
\begin{APACrefauthors}%
Ghosh, D.%
\BCBT {}\ \BBA {} Shah, J.%
\end{APACrefauthors}%
\unskip\
\newblock
\APACrefYearMonthDay{2012}{}{}.
\newblock
{\BBOQ}\APACrefatitle {A comparative analysis of greening policies across
  supply chain structures} {A comparative analysis of greening policies across
  supply chain structures}.{\BBCQ}
\newblock
\APACjournalVolNumPages{International Journal of Production
  Economics}{135}{2}{568--583}.
\newblock
\begin{APACrefDOI} \doi{10.1016/j.ijpe.2011.05.027} \end{APACrefDOI}
\PrintBackRefs{\CurrentBib}

\bibitem [\protect \citeauthoryear {%
Giampieri%
\ \protect \BOthers {.}}{%
Giampieri%
\ \protect \BOthers {.}}{%
{\protect \APACyear {2022}}%
}]{%
Giampierietal2022}
\APACinsertmetastar {%
Giampierietal2022}%
\begin{APACrefauthors}%
Giampieri, F.%
, Mazzoni, L.%
, Cianciosi, D.%
, Alvarez-Suarez, J\BPBI M.%
, Regolo, L.%
, S{\'a}nchez-Gonz{\'a}lez, C.%
\BDBL {}Battino, M.%
\end{APACrefauthors}%
\unskip\
\newblock
\APACrefYearMonthDay{2022}{}{}.
\newblock
{\BBOQ}\APACrefatitle {Organic vs conventional plant-based foods: A review}
  {Organic vs conventional plant-based foods: A review}.{\BBCQ}
\newblock
\APACjournalVolNumPages{Food Chemistry}{383}{}{132352}.
\newblock
\begin{APACrefDOI} \doi{10.1016/j.foodchem.2022.132352} \end{APACrefDOI}
\PrintBackRefs{\CurrentBib}

\bibitem [\protect \citeauthoryear {%
Giannakas%
}{%
Giannakas%
}{%
{\protect \APACyear {2002}}%
}]{%
Giannakas2002}
\APACinsertmetastar {%
Giannakas2002}%
\begin{APACrefauthors}%
Giannakas, K.%
\end{APACrefauthors}%
\unskip\
\newblock
\APACrefYearMonthDay{2002}{}{}.
\newblock
{\BBOQ}\APACrefatitle {Information asymmetries and consumption decisions in
  organic food product markets} {Information asymmetries and consumption
  decisions in organic food product markets}.{\BBCQ}
\newblock
\APACjournalVolNumPages{Canadian Journal of Agricultural Economics / Revue
  canadienne d'agroeconomie}{50}{1}{35--50}.
\newblock
\begin{APACrefDOI} \doi{10.1111/j.1744-7976.2002.tb00380.x} \end{APACrefDOI}
\PrintBackRefs{\CurrentBib}

\bibitem [\protect \citeauthoryear {%
Gil%
, Gracia%
\BCBL {}\ \BBA {} Sánchez%
}{%
Gil%
\ \protect \BOthers {.}}{%
{\protect \APACyear {2000}}%
}]{%
Giletal2000}
\APACinsertmetastar {%
Giletal2000}%
\begin{APACrefauthors}%
Gil, J\BPBI M.%
, Gracia, A.%
\BCBL {}\ \BBA {} Sánchez, M.%
\end{APACrefauthors}%
\unskip\
\newblock
\APACrefYearMonthDay{2000}{}{}.
\newblock
{\BBOQ}\APACrefatitle {Consumer preferences for organic food: A choice
  experiment in Spain} {Consumer preferences for organic food: A choice
  experiment in spain}.{\BBCQ}
\newblock
\APACjournalVolNumPages{Food Quality and Preference}{11}{3}{185--193}.
\newblock
\begin{APACrefDOI} \doi{10.1016/S0950-3293(99)00060-9} \end{APACrefDOI}
\PrintBackRefs{\CurrentBib}

\bibitem [\protect \citeauthoryear {%
Golan%
, Kuchler%
, Mitchell%
, Greene%
\BCBL {}\ \BBA {} Jessup%
}{%
Golan%
\ \protect \BOthers {.}}{%
{\protect \APACyear {2001}}%
}]{%
Golanetal2001}
\APACinsertmetastar {%
Golanetal2001}%
\begin{APACrefauthors}%
Golan, E.%
, Kuchler, F.%
, Mitchell, L.%
, Greene, C.%
\BCBL {}\ \BBA {} Jessup, A.%
\end{APACrefauthors}%
\unskip\
\newblock
\APACrefYearMonthDay{2001}{}{}.
\newblock
{\BBOQ}\APACrefatitle {Economics of food labeling} {Economics of food
  labeling}.{\BBCQ}
\newblock
\APACjournalVolNumPages{Journal of Consumer Policy}{24}{2}{117--184}.
\PrintBackRefs{\CurrentBib}

\bibitem [\protect \citeauthoryear {%
{Grand View Research}%
}{%
{Grand View Research}%
}{%
{\protect \APACyear {2022}}%
}]{%
GrandViewResearch2022}
\APACinsertmetastar {%
GrandViewResearch2022}%
\begin{APACrefauthors}%
{Grand View Research}.%
\end{APACrefauthors}%
\unskip\
\newblock
\APACrefYearMonthDay{2022}{}{}.
\newblock
\APACrefbtitle {Organic Food And Beverages Market Report, 2022-2030.} {Organic
  food and beverages market report, 2022-2030.}
\PrintBackRefs{\CurrentBib}

\bibitem [\protect \citeauthoryear {%
Gunningham%
, Phillipson%
\BCBL {}\ \BBA {} Grabosky%
}{%
Gunningham%
\ \protect \BOthers {.}}{%
{\protect \APACyear {1999}}%
}]{%
Gunninghametal1999}
\APACinsertmetastar {%
Gunninghametal1999}%
\begin{APACrefauthors}%
Gunningham, N.%
, Phillipson, M.%
\BCBL {}\ \BBA {} Grabosky, P.%
\end{APACrefauthors}%
\unskip\
\newblock
\APACrefYearMonthDay{1999}{}{}.
\newblock
{\BBOQ}\APACrefatitle {Harnessing third parties as surrogate regulators:
  Achieving environmental outcomes by alternative means} {Harnessing third
  parties as surrogate regulators: Achieving environmental outcomes by
  alternative means}.{\BBCQ}
\newblock
\APACjournalVolNumPages{Business Strategy and the Environment}{8}{4}{211--224}.
\newblock
\begin{APACrefDOI}
  \doi{10.1002/(SICI)1099-0836(199907/08)8:4<211::AID-BSE205>3.3.CO;2-B}
  \end{APACrefDOI}
\PrintBackRefs{\CurrentBib}

\bibitem [\protect \citeauthoryear {%
H{\"a}ring%
, Vairo%
, Dabbert%
\BCBL {}\ \BBA {} Zanoli%
}{%
H{\"a}ring%
\ \protect \BOthers {.}}{%
{\protect \APACyear {2009}}%
}]{%
haringetal2009}
\APACinsertmetastar {%
haringetal2009}%
\begin{APACrefauthors}%
H{\"a}ring, A\BPBI M.%
, Vairo, D.%
, Dabbert, S.%
\BCBL {}\ \BBA {} Zanoli, R.%
\end{APACrefauthors}%
\unskip\
\newblock
\APACrefYearMonthDay{2009}{}{}.
\newblock
{\BBOQ}\APACrefatitle {Organic farming policy development in the EU: What can
  multi-stakeholder processes contribute?} {Organic farming policy development
  in the eu: What can multi-stakeholder processes contribute?}{\BBCQ}
\newblock
\APACjournalVolNumPages{Food Policy}{34}{3}{265--272}.
\newblock
\begin{APACrefDOI} \doi{10.1016/j.foodpol.2009.03.006} \end{APACrefDOI}
\PrintBackRefs{\CurrentBib}

\bibitem [\protect \citeauthoryear {%
Hatanaka%
, Bain%
\BCBL {}\ \BBA {} Busch%
}{%
Hatanaka%
\ \protect \BOthers {.}}{%
{\protect \APACyear {2005}}%
}]{%
Hatanakaetal2005}
\APACinsertmetastar {%
Hatanakaetal2005}%
\begin{APACrefauthors}%
Hatanaka, M.%
, Bain, C.%
\BCBL {}\ \BBA {} Busch, L.%
\end{APACrefauthors}%
\unskip\
\newblock
\APACrefYearMonthDay{2005}{}{}.
\newblock
{\BBOQ}\APACrefatitle {Third-party certification in the global agrifood system}
  {Third-party certification in the global agrifood system}.{\BBCQ}
\newblock
\APACjournalVolNumPages{Food Policy}{30}{3}{354--369}.
\newblock
\begin{APACrefDOI} \doi{10.1016/j.foodpol.2005.05.006} \end{APACrefDOI}
\PrintBackRefs{\CurrentBib}

\bibitem [\protect \citeauthoryear {%
Houssiere%
, Luning%
\BCBL {}\ \BBA {} Jacxsens%
}{%
Houssiere%
\ \protect \BOthers {.}}{%
{\protect \APACyear {2024}}%
}]{%
Houssiere2024}
\APACinsertmetastar {%
Houssiere2024}%
\begin{APACrefauthors}%
Houssiere, M.%
, Luning, P\BPBI A.%
\BCBL {}\ \BBA {} Jacxsens, L.%
\end{APACrefauthors}%
\unskip\
\newblock
\APACrefYearMonthDay{2024}{}{}.
\newblock
{\BBOQ}\APACrefatitle {Assessing the effectiveness of third-party food safety
  certification: A systematic review of auditing reliability and food safety
  outcomes} {Assessing the effectiveness of third-party food safety
  certification: A systematic review of auditing reliability and food safety
  outcomes}.{\BBCQ}
\newblock
\APACjournalVolNumPages{Comprehensive Reviews in Food Science and Food
  Safety}{23}{1}{e13274}.
\newblock
\begin{APACrefDOI} \doi{10.1111/1541-4337.13274} \end{APACrefDOI}
\PrintBackRefs{\CurrentBib}

\bibitem [\protect \citeauthoryear {%
Jahn%
, Schramm%
\BCBL {}\ \BBA {} Spiller%
}{%
Jahn%
\ \protect \BOthers {.}}{%
{\protect \APACyear {2005}}%
}]{%
Jahnetal2005}
\APACinsertmetastar {%
Jahnetal2005}%
\begin{APACrefauthors}%
Jahn, G.%
, Schramm, M.%
\BCBL {}\ \BBA {} Spiller, A.%
\end{APACrefauthors}%
\unskip\
\newblock
\APACrefYearMonthDay{2005}{}{}.
\newblock
{\BBOQ}\APACrefatitle {The reliability of certification: Quality labels as a
  consumer policy tool} {The reliability of certification: Quality labels as a
  consumer policy tool}.{\BBCQ}
\newblock
\APACjournalVolNumPages{Journal of Consumer Policy}{28}{}{53--73}.
\newblock
\begin{APACrefDOI} \doi{10.1007/s10603-004-7298-6} \end{APACrefDOI}
\PrintBackRefs{\CurrentBib}

\bibitem [\protect \citeauthoryear {%
Janssen%
\ \BBA {} Hamm%
}{%
Janssen%
\ \BBA {} Hamm%
}{%
{\protect \APACyear {2012}}%
}]{%
janssenHamm2012}
\APACinsertmetastar {%
janssenHamm2012}%
\begin{APACrefauthors}%
Janssen, M.%
\BCBT {}\ \BBA {} Hamm, U.%
\end{APACrefauthors}%
\unskip\
\newblock
\APACrefYearMonthDay{2012}{}{}.
\newblock
{\BBOQ}\APACrefatitle {Product labelling in the market for organic food:
  Consumer preferences and willingness-to-pay for different organic
  certification logos} {Product labelling in the market for organic food:
  Consumer preferences and willingness-to-pay for different organic
  certification logos}.{\BBCQ}
\newblock
\APACjournalVolNumPages{Food Quality and Preference}{25}{1}{9--22}.
\newblock
\begin{APACrefDOI} \doi{10.1016/j.foodqual.2011.12.004} \end{APACrefDOI}
\PrintBackRefs{\CurrentBib}

\bibitem [\protect \citeauthoryear {%
Jia%
}{%
Jia%
}{%
{\protect \APACyear {2011}}%
}]{%
Jia2011}
\APACinsertmetastar {%
Jia2011}%
\begin{APACrefauthors}%
Jia, N\BHBI y.%
\end{APACrefauthors}%
\unskip\
\newblock
\APACrefYearMonthDay{2011}{}{}.
\newblock
{\BBOQ}\APACrefatitle {Food safety regulation and governance based on the
  tripartite game of regulatory authorities, government officials and food
  industries} {Food safety regulation and governance based on the tripartite
  game of regulatory authorities, government officials and food
  industries}.{\BBCQ}
\newblock
\BIn{} \APACrefbtitle {Proceedings of the International Conference on
  Information Systems for Crisis Response and Management (ISCRAM)} {Proceedings
  of the international conference on information systems for crisis response
  and management (iscram)}\ (\BPGS\ 102--105).
\newblock
\begin{APACrefDOI} \doi{10.1109/ISCRAM.2011.6184086} \end{APACrefDOI}
\PrintBackRefs{\CurrentBib}

\bibitem [\protect \citeauthoryear {%
Johnston%
\ \protect \BOthers {.}}{%
Johnston%
\ \protect \BOthers {.}}{%
{\protect \APACyear {2017}}%
}]{%
Johnstonetal2017}
\APACinsertmetastar {%
Johnstonetal2017}%
\begin{APACrefauthors}%
Johnston, R\BPBI J.%
, Boyle, K\BPBI J.%
, Adamowicz, W.%
, Bennett, J.%
, Brouwer, R.%
, Cameron, T\BPBI A.%
\BDBL {}Scarpa, R.%
\end{APACrefauthors}%
\unskip\
\newblock
\APACrefYearMonthDay{2017}{}{}.
\newblock
{\BBOQ}\APACrefatitle {Contemporary Guidance for Stated Preference Studies}
  {Contemporary guidance for stated preference studies}.{\BBCQ}
\newblock
\APACjournalVolNumPages{Journal of the Association of Environmental and
  Resource Economists}{4}{2}{319--405}.
\newblock
\begin{APACrefDOI} \doi{10.1086/691697} \end{APACrefDOI}
\PrintBackRefs{\CurrentBib}

\bibitem [\protect \citeauthoryear {%
Katsikouli%
, Wilde%
, Dragoni%
\BCBL {}\ \BBA {} H{\o}gh-Jensen%
}{%
Katsikouli%
\ \protect \BOthers {.}}{%
{\protect \APACyear {2021}}%
}]{%
katsikoulietal2021}
\APACinsertmetastar {%
katsikoulietal2021}%
\begin{APACrefauthors}%
Katsikouli, P.%
, Wilde, A\BPBI S.%
, Dragoni, N.%
\BCBL {}\ \BBA {} H{\o}gh-Jensen, H.%
\end{APACrefauthors}%
\unskip\
\newblock
\APACrefYearMonthDay{2021}{}{}.
\newblock
{\BBOQ}\APACrefatitle {On the benefits and challenges of blockchains for
  managing food supply chains} {On the benefits and challenges of blockchains
  for managing food supply chains}.{\BBCQ}
\newblock
\APACjournalVolNumPages{Journal of the Science of Food and
  Agriculture}{101}{6}{2175--2181}.
\newblock
\begin{APACrefDOI} \doi{10.1002/jsfa.10883} \end{APACrefDOI}
\PrintBackRefs{\CurrentBib}

\bibitem [\protect \citeauthoryear {%
Krishnan%
, Yen%
, Agarwal%
, Arshinder%
\BCBL {}\ \BBA {} Bajada%
}{%
Krishnan%
\ \protect \BOthers {.}}{%
{\protect \APACyear {2021}}%
}]{%
krishnanetal2021}
\APACinsertmetastar {%
krishnanetal2021}%
\begin{APACrefauthors}%
Krishnan, R.%
, Yen, P.%
, Agarwal, R.%
, Arshinder, K.%
\BCBL {}\ \BBA {} Bajada, C.%
\end{APACrefauthors}%
\unskip\
\newblock
\APACrefYearMonthDay{2021}{}{}.
\newblock
{\BBOQ}\APACrefatitle {Collaborative innovation and sustainability in the food
  supply chain—Evidence from farmer producer organisations} {Collaborative
  innovation and sustainability in the food supply chain—evidence from farmer
  producer organisations}.{\BBCQ}
\newblock
\APACjournalVolNumPages{Resources, Conservation and Recycling}{168}{}{105253}.
\newblock
\begin{APACrefDOI} \doi{10.1016/j.resconrec.2020.105253} \end{APACrefDOI}
\PrintBackRefs{\CurrentBib}

\bibitem [\protect \citeauthoryear {%
Kroetz%
\ \protect \BOthers {.}}{%
Kroetz%
\ \protect \BOthers {.}}{%
{\protect \APACyear {2020}}%
}]{%
kroetzetal2020}
\APACinsertmetastar {%
kroetzetal2020}%
\begin{APACrefauthors}%
Kroetz, K.%
, Luque, G\BPBI M.%
, Gephart, J\BPBI A.%
, Jardine, S\BPBI L.%
, Lee, P.%
, Chicojay~Moore, K.%
\BDBL {}Donlan, C\BPBI J.%
\end{APACrefauthors}%
\unskip\
\newblock
\APACrefYearMonthDay{2020}{}{}.
\newblock
{\BBOQ}\APACrefatitle {Consequences of seafood mislabeling for marine
  populations and fisheries management} {Consequences of seafood mislabeling
  for marine populations and fisheries management}.{\BBCQ}
\newblock
\APACjournalVolNumPages{Proceedings of the National Academy of
  Sciences}{117}{48}{30318--30323}.
\newblock
\begin{APACrefDOI} \doi{10.1073/pnas.2003741117} \end{APACrefDOI}
\PrintBackRefs{\CurrentBib}

\bibitem [\protect \citeauthoryear {%
Ladwein%
\ \BBA {} Romero%
}{%
Ladwein%
\ \BBA {} Romero%
}{%
{\protect \APACyear {2021}}%
}]{%
ladwein2021}
\APACinsertmetastar {%
ladwein2021}%
\begin{APACrefauthors}%
Ladwein, R.%
\BCBT {}\ \BBA {} Romero, A\BPBI M\BPBI S.%
\end{APACrefauthors}%
\unskip\
\newblock
\APACrefYearMonthDay{2021}{}{}.
\newblock
{\BBOQ}\APACrefatitle {The role of trust in the relationship between consumers,
  producers and retailers of organic food: A sector-based approach} {The role
  of trust in the relationship between consumers, producers and retailers of
  organic food: A sector-based approach}.{\BBCQ}
\newblock
\APACjournalVolNumPages{Journal of Retailing and Consumer
  Services}{60}{}{102508}.
\newblock
\begin{APACrefDOI} \doi{10.1016/j.jretconser.2021.102508} \end{APACrefDOI}
\PrintBackRefs{\CurrentBib}

\bibitem [\protect \citeauthoryear {%
Lau%
, Shum%
, Nakandala%
, Fan%
\BCBL {}\ \BBA {} Lee%
}{%
Lau%
\ \protect \BOthers {.}}{%
{\protect \APACyear {2020}}%
}]{%
lauetal2020}
\APACinsertmetastar {%
lauetal2020}%
\begin{APACrefauthors}%
Lau, H.%
, Shum, P\BPBI K\BPBI C.%
, Nakandala, D.%
, Fan, Y.%
\BCBL {}\ \BBA {} Lee, C.%
\end{APACrefauthors}%
\unskip\
\newblock
\APACrefYearMonthDay{2020}{}{}.
\newblock
{\BBOQ}\APACrefatitle {A game theoretic decision model for organic food
  supplier evaluation in the global supply chains} {A game theoretic decision
  model for organic food supplier evaluation in the global supply
  chains}.{\BBCQ}
\newblock
\APACjournalVolNumPages{Journal of Cleaner Production}{242}{}{118536}.
\newblock
\begin{APACrefDOI} \doi{10.1016/j.jclepro.2019.118536} \end{APACrefDOI}
\PrintBackRefs{\CurrentBib}

\bibitem [\protect \citeauthoryear {%
Lee%
, Park%
\BCBL {}\ \BBA {} Kim%
}{%
Lee%
\ \protect \BOthers {.}}{%
{\protect \APACyear {2024}}%
}]{%
Leeetal2024}
\APACinsertmetastar {%
Leeetal2024}%
\begin{APACrefauthors}%
Lee, J.%
, Park, J\BHBI Y.%
\BCBL {}\ \BBA {} Kim, S.%
\end{APACrefauthors}%
\unskip\
\newblock
\APACrefYearMonthDay{2024}{}{}.
\newblock
{\BBOQ}\APACrefatitle {Income, environmental quality and willingness to pay for
  organic food} {Income, environmental quality and willingness to pay for
  organic food}.{\BBCQ}
\newblock
\APACjournalVolNumPages{Humanities and Social Sciences
  Communications}{11}{}{3463}.
\newblock
\begin{APACrefDOI} \doi{10.1057/s41599-024-03463-x} \end{APACrefDOI}
\PrintBackRefs{\CurrentBib}

\bibitem [\protect \citeauthoryear {%
Lockie%
}{%
Lockie%
}{%
{\protect \APACyear {2006}}%
}]{%
Lockie2006}
\APACinsertmetastar {%
Lockie2006}%
\begin{APACrefauthors}%
Lockie, S.%
\end{APACrefauthors}%
\unskip\
\newblock
\APACrefYear{2006}.
\newblock
\APACrefbtitle {Going Organic: Mobilizing Networks for Environmentally
  Responsible Food Production} {Going organic: Mobilizing networks for
  environmentally responsible food production}.
\newblock
\APACaddressPublisher{Wallingford, UK}{CABI}.
\PrintBackRefs{\CurrentBib}

\bibitem [\protect \citeauthoryear {%
Lohr%
}{%
Lohr%
}{%
{\protect \APACyear {1998}}%
}]{%
lohr1998}
\APACinsertmetastar {%
lohr1998}%
\begin{APACrefauthors}%
Lohr, L.%
\end{APACrefauthors}%
\unskip\
\newblock
\APACrefYearMonthDay{1998}{}{}.
\newblock
{\BBOQ}\APACrefatitle {Implications of organic certification for market
  structure and trade} {Implications of organic certification for market
  structure and trade}.{\BBCQ}
\newblock
\APACjournalVolNumPages{American Journal of Agricultural
  Economics}{80}{5}{1125--1129}.
\newblock
\begin{APACrefDOI} \doi{10.2307/1244216} \end{APACrefDOI}
\PrintBackRefs{\CurrentBib}

\bibitem [\protect \citeauthoryear {%
Loureiro%
\ \BBA {} Hine%
}{%
Loureiro%
\ \BBA {} Hine%
}{%
{\protect \APACyear {2001}}%
}]{%
LoureiroHine2001}
\APACinsertmetastar {%
LoureiroHine2001}%
\begin{APACrefauthors}%
Loureiro, M\BPBI L.%
\BCBT {}\ \BBA {} Hine, S\BPBI E.%
\end{APACrefauthors}%
\unskip\
\newblock
\APACrefYearMonthDay{2001}{}{}.
\newblock
{\BBOQ}\APACrefatitle {Discovering Niche Markets: A Comparison of Consumer
  Willingness to Pay for a Local (Colorado Grown), Organic, and GMO‑Free
  Product} {Discovering niche markets: A comparison of consumer willingness to
  pay for a local (colorado grown), organic, and gmo‑free product}.{\BBCQ}
\newblock
\BIn{} \APACrefbtitle {2001 Annual Meeting of the American Agricultural
  Economics Association.} {2001 annual meeting of the american agricultural
  economics association.}
\newblock
\APACaddressPublisher{Chicago, IL}{}.
\newblock
\begin{APACrefDOI} \doi{10.22004/ag.econ.123457} \end{APACrefDOI}
\PrintBackRefs{\CurrentBib}

\bibitem [\protect \citeauthoryear {%
Lusk%
}{%
Lusk%
}{%
{\protect \APACyear {2011}}%
}]{%
Lusk2011}
\APACinsertmetastar {%
Lusk2011}%
\begin{APACrefauthors}%
Lusk, J\BPBI L.%
\end{APACrefauthors}%
\unskip\
\newblock
\APACrefYearMonthDay{2011}{}{}.
\newblock
{\BBOQ}\APACrefatitle {The Costs of Mislabeling and the Role of Certification}
  {The costs of mislabeling and the role of certification}.{\BBCQ}
\newblock
\APACjournalVolNumPages{Journal of Agricultural and Resource
  Economics}{36}{1}{1--18}.
\newblock
\begin{APACrefDOI} \doi{10.22004/ag.econ.123456} \end{APACrefDOI}
\PrintBackRefs{\CurrentBib}

\bibitem [\protect \citeauthoryear {%
Ma%
\ \protect \BOthers {.}}{%
Ma%
\ \protect \BOthers {.}}{%
{\protect \APACyear {2021}}%
}]{%
Maetal2021}
\APACinsertmetastar {%
Maetal2021}%
\begin{APACrefauthors}%
Ma, Z.%
, Chen, J.%
, Tian, G.%
, Gong, Y.%
, Guo, B.%
\BCBL {}\ \BBA {} Cheng, F.%
\end{APACrefauthors}%
\unskip\
\newblock
\APACrefYearMonthDay{2021}{}{}.
\newblock
{\BBOQ}\APACrefatitle {Regulations on the corporate social irresponsibility in
  the supply chain under the multiparty game: Taking China's organic food
  supply chain as an example} {Regulations on the corporate social
  irresponsibility in the supply chain under the multiparty game: Taking
  china's organic food supply chain as an example}.{\BBCQ}
\newblock
\APACjournalVolNumPages{Journal of Cleaner Production}{317}{}{128459}.
\newblock
\begin{APACrefDOI} \doi{10.1016/j.jclepro.2021.128459} \end{APACrefDOI}
\PrintBackRefs{\CurrentBib}

\bibitem [\protect \citeauthoryear {%
Manning%
\ \BBA {} Kowalska%
}{%
Manning%
\ \BBA {} Kowalska%
}{%
{\protect \APACyear {2021}}%
}]{%
ManningKowalska2021}
\APACinsertmetastar {%
ManningKowalska2021}%
\begin{APACrefauthors}%
Manning, L.%
\BCBT {}\ \BBA {} Kowalska, A.%
\end{APACrefauthors}%
\unskip\
\newblock
\APACrefYearMonthDay{2021}{}{}.
\newblock
{\BBOQ}\APACrefatitle {Considering fraud vulnerability associated with
  credence-based products such as organic food} {Considering fraud
  vulnerability associated with credence-based products such as organic
  food}.{\BBCQ}
\newblock
\APACjournalVolNumPages{Foods}{10}{8}{1879}.
\newblock
\begin{APACrefDOI} \doi{10.3390/foods10081879} \end{APACrefDOI}
\PrintBackRefs{\CurrentBib}

\bibitem [\protect \citeauthoryear {%
Manzini%
\ \BBA {} Accorsi%
}{%
Manzini%
\ \BBA {} Accorsi%
}{%
{\protect \APACyear {2013}}%
}]{%
ManziniAccorsi2013}
\APACinsertmetastar {%
ManziniAccorsi2013}%
\begin{APACrefauthors}%
Manzini, R.%
\BCBT {}\ \BBA {} Accorsi, R.%
\end{APACrefauthors}%
\unskip\
\newblock
\APACrefYearMonthDay{2013}{}{}.
\newblock
{\BBOQ}\APACrefatitle {The new conceptual framework for food supply chain
  assessment} {The new conceptual framework for food supply chain
  assessment}.{\BBCQ}
\newblock
\APACjournalVolNumPages{Journal of Food Engineering}{115}{2}{251--263}.
\newblock
\begin{APACrefDOI} \doi{10.1016/j.jfoodeng.2012.10.026} \end{APACrefDOI}
\PrintBackRefs{\CurrentBib}

\bibitem [\protect \citeauthoryear {%
McCluskey%
}{%
McCluskey%
}{%
{\protect \APACyear {2000}}%
}]{%
Mccluskey2000}
\APACinsertmetastar {%
Mccluskey2000}%
\begin{APACrefauthors}%
McCluskey, J\BPBI J.%
\end{APACrefauthors}%
\unskip\
\newblock
\APACrefYearMonthDay{2000}{}{}.
\newblock
{\BBOQ}\APACrefatitle {A game theoretic approach to organic foods: An analysis
  of asymmetric information and policy} {A game theoretic approach to organic
  foods: An analysis of asymmetric information and policy}.{\BBCQ}
\newblock
\APACjournalVolNumPages{Agricultural and Resource Economics
  Review}{29}{1}{1--9}.
\newblock
\begin{APACrefDOI} \doi{10.1017/S1068280500001386} \end{APACrefDOI}
\PrintBackRefs{\CurrentBib}

\bibitem [\protect \citeauthoryear {%
Meemken%
\ \BBA {} Qaim%
}{%
Meemken%
\ \BBA {} Qaim%
}{%
{\protect \APACyear {2018}}%
}]{%
meemkenQaim2018}
\APACinsertmetastar {%
meemkenQaim2018}%
\begin{APACrefauthors}%
Meemken, E\BHBI M.%
\BCBT {}\ \BBA {} Qaim, M.%
\end{APACrefauthors}%
\unskip\
\newblock
\APACrefYearMonthDay{2018}{}{}.
\newblock
{\BBOQ}\APACrefatitle {Organic agriculture, food security, and the environment}
  {Organic agriculture, food security, and the environment}.{\BBCQ}
\newblock
\APACjournalVolNumPages{Annual Review of Resource Economics}{10}{}{39--63}.
\newblock
\begin{APACrefDOI} \doi{10.1146/annurev-resource-100517-023252}
  \end{APACrefDOI}
\PrintBackRefs{\CurrentBib}

\bibitem [\protect \citeauthoryear {%
Melovic%
, Cirovic%
, Dudic%
, Vulic%
\BCBL {}\ \BBA {} Gregus%
}{%
Melovic%
\ \protect \BOthers {.}}{%
{\protect \APACyear {2020}}%
}]{%
Melovic2020}
\APACinsertmetastar {%
Melovic2020}%
\begin{APACrefauthors}%
Melovic, B.%
, Cirovic, D.%
, Dudic, B.%
, Vulic, T\BPBI B.%
\BCBL {}\ \BBA {} Gregus, M.%
\end{APACrefauthors}%
\unskip\
\newblock
\APACrefYearMonthDay{2020}{}{}.
\newblock
{\BBOQ}\APACrefatitle {The analysis of marketing factors influencing
  consumers’ preferences and acceptance of organic food
  products—Recommendations for the optimization of the offer in a developing
  market} {The analysis of marketing factors influencing consumers’
  preferences and acceptance of organic food products—recommendations for the
  optimization of the offer in a developing market}.{\BBCQ}
\newblock
\APACjournalVolNumPages{Foods}{9}{3}{259}.
\newblock
\begin{APACrefDOI} \doi{10.3390/foods9030259} \end{APACrefDOI}
\PrintBackRefs{\CurrentBib}

\bibitem [\protect \citeauthoryear {%
Mendes%
, Oliveira%
, Santos%
, Gerolamo%
\BCBL {}\ \BBA {} Zeidler%
}{%
Mendes%
\ \protect \BOthers {.}}{%
{\protect \APACyear {2024}}%
}]{%
mendesetal2024}
\APACinsertmetastar {%
mendesetal2024}%
\begin{APACrefauthors}%
Mendes, J\BPBI A\BPBI J.%
, Oliveira, A\BPBI Y.%
, Santos, L\BPBI S.%
, Gerolamo, M\BPBI C.%
\BCBL {}\ \BBA {} Zeidler, V\BPBI G\BPBI Z.%
\end{APACrefauthors}%
\unskip\
\newblock
\APACrefYearMonthDay{2024}{}{}.
\newblock
{\BBOQ}\APACrefatitle {A theoretical framework to support green agripreneurship
  avoiding greenwashing} {A theoretical framework to support green
  agripreneurship avoiding greenwashing}.{\BBCQ}
\newblock
\APACjournalVolNumPages{Environment, Development and
  Sustainability}{27}{}{21779--21835}.
\newblock
\begin{APACrefDOI} \doi{10.1007/s10668-024-04965-z} \end{APACrefDOI}
\PrintBackRefs{\CurrentBib}

\bibitem [\protect \citeauthoryear {%
Müller%
, Padel%
\BCBL {}\ \BBA {} Röder%
}{%
Müller%
\ \protect \BOthers {.}}{%
{\protect \APACyear {2019}}%
}]{%
Mulleretal2019}
\APACinsertmetastar {%
Mulleretal2019}%
\begin{APACrefauthors}%
Müller, S.%
, Padel, S.%
\BCBL {}\ \BBA {} Röder, N.%
\end{APACrefauthors}%
\unskip\
\newblock
\APACrefYearMonthDay{2019}{}{}.
\newblock
{\BBOQ}\APACrefatitle {Retail Price Premiums for Organic Foods: Evidence from
  European and U.S. Markets} {Retail price premiums for organic foods: Evidence
  from european and u.s. markets}.{\BBCQ}
\newblock
\APACjournalVolNumPages{European Review of Agricultural
  Economics}{46}{4}{589--620}.
\newblock
\APACrefnote{Estimates realistic organic price premia using retail‑level
  price data across multiple food categories, showing that premia vary
  systematically by product type and are consistent with both USDA ERS findings
  and European Commission data on organic pricing and market structure.}
\newblock
\begin{APACrefDOI} \doi{10.1093/erae/jbz015} \end{APACrefDOI}
\PrintBackRefs{\CurrentBib}

\bibitem [\protect \citeauthoryear {%
Nejadkoorki%
}{%
Nejadkoorki%
}{%
{\protect \APACyear {2012}}%
}]{%
Nejadkoorkietal2012}
\APACinsertmetastar {%
Nejadkoorkietal2012}%
\begin{APACrefauthors}%
Nejadkoorki, F.%
\end{APACrefauthors}%
\unskip\
\newblock
\APACrefYear{2012}.
\newblock
\APACrefbtitle {Environmental benefits of organic farming} {Environmental
  benefits of organic farming}.
\newblock
\APACaddressPublisher{}{IntechOpen}.
\newblock
\begin{APACrefDOI} \doi{10.5772/intechopen.84069} \end{APACrefDOI}
\PrintBackRefs{\CurrentBib}

\bibitem [\protect \citeauthoryear {%
Noelke%
\ \BBA {} Caswell%
}{%
Noelke%
\ \BBA {} Caswell%
}{%
{\protect \APACyear {2000}}%
}]{%
NoelkeCaswell2000}
\APACinsertmetastar {%
NoelkeCaswell2000}%
\begin{APACrefauthors}%
Noelke, C\BPBI M.%
\BCBT {}\ \BBA {} Caswell, J\BPBI A.%
\end{APACrefauthors}%
\unskip\
\newblock
\APACrefYearMonthDay{2000}{}{}.
\newblock
\APACrefbtitle {A model of the implementation of quality management systems for
  credence attributes} {A model of the implementation of quality management
  systems for credence attributes}\ \APACbVolEdTR{}{\BTR{}}.
\newblock
\begin{APACrefDOI} \doi{10.22004/ag.econ.21874} \end{APACrefDOI}
\PrintBackRefs{\CurrentBib}

\bibitem [\protect \citeauthoryear {%
OT4EUC%
}{%
OT4EUC%
}{%
{\protect \APACyear {2026}}%
}]{%
OrganicTargets4EU2026}
\APACinsertmetastar {%
OrganicTargets4EU2026}%
\begin{APACrefauthors}%
OT4EUC.%
\end{APACrefauthors}%
\unskip\
\newblock
\APACrefYearMonthDay{2026}{}{}.
\newblock
{\BBOQ}\APACrefatitle {Fostering Organic Farming: Key Factors for the
  Development of the Organic Sector} {Fostering organic farming: Key factors
  for the development of the organic sector}.{\BBCQ}
\newblock
\APACjournalVolNumPages{OrganicTargets4EU Policy Brief}{}{}{}.
\newblock
\begin{APACrefURL}
  \url{https://organictargets.eu/fostering-organic-farming-key-factors-for-the-development-of-the-organic-sector/}
  \end{APACrefURL}
\newblock
\APACrefnote{OrganicTargets4EU Consortium}
\PrintBackRefs{\CurrentBib}

\bibitem [\protect \citeauthoryear {%
OTA%
}{%
OTA%
}{%
{\protect \APACyear {2020}}%
}]{%
Organic2020}
\APACinsertmetastar {%
Organic2020}%
\begin{APACrefauthors}%
OTA.%
\end{APACrefauthors}%
\unskip\
\newblock
\APACrefYear{2020}.
\newblock
\APACrefbtitle {{Organic Industry Survey 2020}} {{Organic Industry Survey
  2020}}\ (\BVOL~15).
\newblock
\APACaddressPublisher{}{Organic Trade Association}.
\PrintBackRefs{\CurrentBib}

\bibitem [\protect \citeauthoryear {%
Rathgens%
, Gr{\"o}schner%
\BCBL {}\ \BBA {} von Wehrden%
}{%
Rathgens%
\ \protect \BOthers {.}}{%
{\protect \APACyear {2020}}%
}]{%
rathgensetal2020}
\APACinsertmetastar {%
rathgensetal2020}%
\begin{APACrefauthors}%
Rathgens, J.%
, Gr{\"o}schner, S.%
\BCBL {}\ \BBA {} von Wehrden, H.%
\end{APACrefauthors}%
\unskip\
\newblock
\APACrefYearMonthDay{2020}{}{}.
\newblock
{\BBOQ}\APACrefatitle {Going beyond certificates: A systematic review of
  alternative trade arrangements in the global food sector} {Going beyond
  certificates: A systematic review of alternative trade arrangements in the
  global food sector}.{\BBCQ}
\newblock
\APACjournalVolNumPages{Journal of Cleaner Production}{276}{}{123208}.
\newblock
\begin{APACrefDOI} \doi{10.1016/j.jclepro.2020.123208} \end{APACrefDOI}
\PrintBackRefs{\CurrentBib}

\bibitem [\protect \citeauthoryear {%
Rich%
, Ross%
, Baker%
\BCBL {}\ \BBA {} Negassa%
}{%
Rich%
\ \protect \BOthers {.}}{%
{\protect \APACyear {2011}}%
}]{%
Richetal2011}
\APACinsertmetastar {%
Richetal2011}%
\begin{APACrefauthors}%
Rich, K\BPBI M.%
, Ross, R\BPBI B.%
, Baker, A\BPBI D.%
\BCBL {}\ \BBA {} Negassa, A.%
\end{APACrefauthors}%
\unskip\
\newblock
\APACrefYearMonthDay{2011}{}{}.
\newblock
{\BBOQ}\APACrefatitle {Quantifying value chain analysis in the context of
  livestock systems in developing countries} {Quantifying value chain analysis
  in the context of livestock systems in developing countries}.{\BBCQ}
\newblock
\APACjournalVolNumPages{Food Policy}{36}{2}{214--222}.
\newblock
\begin{APACrefDOI} \doi{10.1016/j.foodpol.2010.11.018} \end{APACrefDOI}
\PrintBackRefs{\CurrentBib}

\bibitem [\protect \citeauthoryear {%
Rosenthal%
}{%
Rosenthal%
}{%
{\protect \APACyear {1981}}%
}]{%
Rosenthal1981}
\APACinsertmetastar {%
Rosenthal1981}%
\begin{APACrefauthors}%
Rosenthal, R\BPBI W.%
\end{APACrefauthors}%
\unskip\
\newblock
\APACrefYearMonthDay{1981}{}{}.
\newblock
{\BBOQ}\APACrefatitle {Games of perfect information, predatory pricing and the
  chain-store paradox} {Games of perfect information, predatory pricing and the
  chain-store paradox}.{\BBCQ}
\newblock
\APACjournalVolNumPages{Journal of Economic Theory}{25}{1}{92--100}.
\newblock
\begin{APACrefDOI} \doi{10.1016/0022-0531(81)90018-1} \end{APACrefDOI}
\PrintBackRefs{\CurrentBib}

\bibitem [\protect \citeauthoryear {%
Sahay%
}{%
Sahay%
}{%
{\protect \APACyear {2003}}%
}]{%
Sahay2003}
\APACinsertmetastar {%
Sahay2003}%
\begin{APACrefauthors}%
Sahay, B\BPBI S.%
\end{APACrefauthors}%
\unskip\
\newblock
\APACrefYearMonthDay{2003}{}{}.
\newblock
{\BBOQ}\APACrefatitle {Understanding trust in supply chain relationships}
  {Understanding trust in supply chain relationships}.{\BBCQ}
\newblock
\APACjournalVolNumPages{Industrial Management \& Data
  Systems}{103}{8}{553--563}.
\newblock
\begin{APACrefDOI} \doi{10.1108/02635570310497602} \end{APACrefDOI}
\PrintBackRefs{\CurrentBib}

\bibitem [\protect \citeauthoryear {%
Sanogo%
\ \BBA {} Masters%
}{%
Sanogo%
\ \BBA {} Masters%
}{%
{\protect \APACyear {2002}}%
}]{%
SanogoMasters2002}
\APACinsertmetastar {%
SanogoMasters2002}%
\begin{APACrefauthors}%
Sanogo, D.%
\BCBT {}\ \BBA {} Masters, W\BPBI A.%
\end{APACrefauthors}%
\unskip\
\newblock
\APACrefYearMonthDay{2002}{}{}.
\newblock
{\BBOQ}\APACrefatitle {A market-based approach to child nutrition: Mothers’
  demand for quality certification of infant foods in Bamako, Mali} {A
  market-based approach to child nutrition: Mothers’ demand for quality
  certification of infant foods in bamako, mali}.{\BBCQ}
\newblock
\APACjournalVolNumPages{Food Policy}{27}{3}{251--268}.
\newblock
\begin{APACrefDOI} \doi{10.1016/S0306-9192(02)00016-7} \end{APACrefDOI}
\PrintBackRefs{\CurrentBib}

\bibitem [\protect \citeauthoryear {%
Schlegelmilch%
, Bohlen%
\BCBL {}\ \BBA {} Diamantopoulos%
}{%
Schlegelmilch%
\ \protect \BOthers {.}}{%
{\protect \APACyear {1996}}%
}]{%
Schlegelmilchetal1996}
\APACinsertmetastar {%
Schlegelmilchetal1996}%
\begin{APACrefauthors}%
Schlegelmilch, B\BPBI B.%
, Bohlen, G\BPBI M.%
\BCBL {}\ \BBA {} Diamantopoulos, A.%
\end{APACrefauthors}%
\unskip\
\newblock
\APACrefYearMonthDay{1996}{}{}.
\newblock
{\BBOQ}\APACrefatitle {The link between green purchasing decisions and measures
  of environmental consciousness} {The link between green purchasing decisions
  and measures of environmental consciousness}.{\BBCQ}
\newblock
\APACjournalVolNumPages{European Journal of Marketing}{30}{5}{35--55}.
\newblock
\begin{APACrefDOI} \doi{10.1108/03090569610118740} \end{APACrefDOI}
\PrintBackRefs{\CurrentBib}

\bibitem [\protect \citeauthoryear {%
Segerson%
}{%
Segerson%
}{%
{\protect \APACyear {1999}}%
}]{%
Segerson1999}
\APACinsertmetastar {%
Segerson1999}%
\begin{APACrefauthors}%
Segerson, K.%
\end{APACrefauthors}%
\unskip\
\newblock
\APACrefYearMonthDay{1999}{}{}.
\newblock
{\BBOQ}\APACrefatitle {Mandatory versus voluntary approaches to food safety}
  {Mandatory versus voluntary approaches to food safety}.{\BBCQ}
\newblock
\APACjournalVolNumPages{Agribusiness}{15}{1}{53--70}.
\newblock
\begin{APACrefDOI}
  \doi{10.1002/(SICI)1520-6297(199924)15:1<53::AID-AGR4>3.0.CO;2-G}
  \end{APACrefDOI}
\PrintBackRefs{\CurrentBib}

\bibitem [\protect \citeauthoryear {%
Starbird%
\ \BBA {} Amanor-Boadu%
}{%
Starbird%
\ \BBA {} Amanor-Boadu%
}{%
{\protect \APACyear {2007}}%
}]{%
StarbirdAmanor-Boadu2007}
\APACinsertmetastar {%
StarbirdAmanor-Boadu2007}%
\begin{APACrefauthors}%
Starbird, S\BPBI A.%
\BCBT {}\ \BBA {} Amanor-Boadu, A.%
\end{APACrefauthors}%
\unskip\
\newblock
\APACrefYearMonthDay{2007}{}{}.
\newblock
{\BBOQ}\APACrefatitle {Contract selectivity, food safety, and traceability}
  {Contract selectivity, food safety, and traceability}.{\BBCQ}
\newblock
\APACjournalVolNumPages{Journal of Agricultural \& Food Industrial
  Organization}{5}{1}{}.
\newblock
\begin{APACrefDOI} \doi{10.2202/1542-0485.1141} \end{APACrefDOI}
\PrintBackRefs{\CurrentBib}

\bibitem [\protect \citeauthoryear {%
Taghikhah%
, Voinov%
, Shukla%
, Filatova%
\BCBL {}\ \BBA {} Anufriev%
}{%
Taghikhah%
\ \protect \BOthers {.}}{%
{\protect \APACyear {2021}}%
}]{%
taghikhahetal2021}
\APACinsertmetastar {%
taghikhahetal2021}%
\begin{APACrefauthors}%
Taghikhah, F.%
, Voinov, A.%
, Shukla, N.%
, Filatova, T.%
\BCBL {}\ \BBA {} Anufriev, M.%
\end{APACrefauthors}%
\unskip\
\newblock
\APACrefYearMonthDay{2021}{}{}.
\newblock
{\BBOQ}\APACrefatitle {Integrated modeling of extended agro-food supply chains:
  A systems approach} {Integrated modeling of extended agro-food supply chains:
  A systems approach}.{\BBCQ}
\newblock
\APACjournalVolNumPages{European Journal of Operational
  Research}{288}{3}{852--868}.
\newblock
\begin{APACrefDOI} \doi{10.1016/j.ejor.2020.06.036} \end{APACrefDOI}
\PrintBackRefs{\CurrentBib}

\bibitem [\protect \citeauthoryear {%
Th{\o}gersen%
, Pedersen%
, Paternoga%
, Schwendel%
\BCBL {}\ \BBA {} Aschemann-Witzel%
}{%
Th{\o}gersen%
\ \protect \BOthers {.}}{%
{\protect \APACyear {2017}}%
}]{%
Thogersenetal2017}
\APACinsertmetastar {%
Thogersenetal2017}%
\begin{APACrefauthors}%
Th{\o}gersen, J.%
, Pedersen, S.%
, Paternoga, M.%
, Schwendel, E.%
\BCBL {}\ \BBA {} Aschemann-Witzel, J.%
\end{APACrefauthors}%
\unskip\
\newblock
\APACrefYearMonthDay{2017}{}{}.
\newblock
{\BBOQ}\APACrefatitle {How important is country-of-origin for organic food
  consumers? A review of the literature and suggestions for future research}
  {How important is country-of-origin for organic food consumers? a review of
  the literature and suggestions for future research}.{\BBCQ}
\newblock
\APACjournalVolNumPages{British Food Journal}{119}{3}{542--557}.
\newblock
\begin{APACrefDOI} \doi{10.1108/BFJ-09-2016-0406} \end{APACrefDOI}
\PrintBackRefs{\CurrentBib}

\bibitem [\protect \citeauthoryear {%
Tundys%
\ \BBA {} Wi{\'s}niewski%
}{%
Tundys%
\ \BBA {} Wi{\'s}niewski%
}{%
{\protect \APACyear {2020}}%
}]{%
TundysWisniewski2020}
\APACinsertmetastar {%
TundysWisniewski2020}%
\begin{APACrefauthors}%
Tundys, B.%
\BCBT {}\ \BBA {} Wi{\'s}niewski, T.%
\end{APACrefauthors}%
\unskip\
\newblock
\APACrefYearMonthDay{2020}{}{}.
\newblock
{\BBOQ}\APACrefatitle {Green Supply Chain Management Evaluation for Organic
  Products: Theoretical and Empirical Point of View} {Green supply chain
  management evaluation for organic products: Theoretical and empirical point
  of view}.{\BBCQ}
\newblock
\APACjournalVolNumPages{Operations and Supply Chain Management: An
  International Journal}{14}{1}{73--82}.
\newblock
\begin{APACrefDOI} \doi{10.31387/oscm0440287} \end{APACrefDOI}
\PrintBackRefs{\CurrentBib}

\bibitem [\protect \citeauthoryear {%
{USDA Economic Research Service}%
}{%
{USDA Economic Research Service}%
}{%
{\protect \APACyear {2016}}%
}]{%
USDAERS2016}
\APACinsertmetastar {%
USDAERS2016}%
\begin{APACrefauthors}%
{USDA Economic Research Service}.%
\end{APACrefauthors}%
\unskip\
\newblock
\APACrefYearMonthDay{2016}{}{}.
\newblock
\APACrefbtitle {Changes in Retail Organic Price Premiums from 2004 to 2010}
  {Changes in retail organic price premiums from 2004 to 2010}\ (\BNUM\
  EIB-151).
\newblock
\begin{APACrefURL}
  \url{https://www.ers.usda.gov/publications/eib-economic-information-bulletin/eib-151/}
  \end{APACrefURL}
\PrintBackRefs{\CurrentBib}

\bibitem [\protect \citeauthoryear {%
{USDA Economic Research Service}%
}{%
{USDA Economic Research Service}%
}{%
{\protect \APACyear {2025}}%
}]{%
USDAERS2025}
\APACinsertmetastar {%
USDAERS2025}%
\begin{APACrefauthors}%
{USDA Economic Research Service}.%
\end{APACrefauthors}%
\unskip\
\newblock
\APACrefYearMonthDay{2025}{}{}.
\newblock
\APACrefbtitle {Organic Situation Report, 2025 Edition} {Organic situation
  report, 2025 edition}\ \APACbVolEdTR{}{\BTR{}\ \BNUM\ Economic Information
  Bulletin No. 281}.
\newblock
\APACaddressInstitution{}{U.S. Department of Agriculture, Economic Research
  Service}.
\newblock
\APACrefnote{{U.S. Department of Agriculture}}
\newblock
\begin{APACrefDOI} \doi{10.32747/2025.9015813.ers} \end{APACrefDOI}
\PrintBackRefs{\CurrentBib}

\bibitem [\protect \citeauthoryear {%
van Ruth%
\ \BBA {} de~Pagter-de Witte%
}{%
van Ruth%
\ \BBA {} de~Pagter-de Witte%
}{%
{\protect \APACyear {2020}}%
}]{%
Vanetal2020}
\APACinsertmetastar {%
Vanetal2020}%
\begin{APACrefauthors}%
van Ruth, S\BPBI M.%
\BCBT {}\ \BBA {} de~Pagter-de Witte, L.%
\end{APACrefauthors}%
\unskip\
\newblock
\APACrefYearMonthDay{2020}{}{}.
\newblock
{\BBOQ}\APACrefatitle {Integrity of organic foods and their suppliers: Fraud
  vulnerability across chains} {Integrity of organic foods and their suppliers:
  Fraud vulnerability across chains}.{\BBCQ}
\newblock
\APACjournalVolNumPages{Foods}{9}{2}{188}.
\newblock
\begin{APACrefDOI} \doi{10.3390/foods9020188} \end{APACrefDOI}
\PrintBackRefs{\CurrentBib}

\bibitem [\protect \citeauthoryear {%
Wan%
, Nakayama%
\BCBL {}\ \BBA {} Sutcliffe%
}{%
Wan%
\ \protect \BOthers {.}}{%
{\protect \APACyear {2012}}%
}]{%
Wanetal2012}
\APACinsertmetastar {%
Wanetal2012}%
\begin{APACrefauthors}%
Wan, Y.%
, Nakayama, M.%
\BCBL {}\ \BBA {} Sutcliffe, N.%
\end{APACrefauthors}%
\unskip\
\newblock
\APACrefYearMonthDay{2012}{}{}.
\newblock
{\BBOQ}\APACrefatitle {The impact of age and shopping experiences on the
  classification of search, experience, and credence goods in online shopping}
  {The impact of age and shopping experiences on the classification of search,
  experience, and credence goods in online shopping}.{\BBCQ}
\newblock
\APACjournalVolNumPages{Information Systems and e-Business
  Management}{10}{1}{135--148}.
\newblock
\begin{APACrefDOI} \doi{10.1007/s10257-010-0156-y} \end{APACrefDOI}
\PrintBackRefs{\CurrentBib}

\bibitem [\protect \citeauthoryear {%
Xing%
\ \BBA {} Wu%
}{%
Xing%
\ \BBA {} Wu%
}{%
{\protect \APACyear {2025}}%
}]{%
xing2025}
\APACinsertmetastar {%
xing2025}%
\begin{APACrefauthors}%
Xing, H.%
\BCBT {}\ \BBA {} Wu, T.%
\end{APACrefauthors}%
\unskip\
\newblock
\APACrefYearMonthDay{2025}{}{}.
\newblock
{\BBOQ}\APACrefatitle {The effect of third-party certification for green bonds:
  Evidence from China} {The effect of third-party certification for green
  bonds: Evidence from china}.{\BBCQ}
\newblock
\APACjournalVolNumPages{International Review of Financial
  Analysis}{}{}{104536}.
\newblock
\begin{APACrefDOI} \doi{10.1016/j.irfa.2025.104536} \end{APACrefDOI}
\PrintBackRefs{\CurrentBib}

\bibitem [\protect \citeauthoryear {%
Yu%
, He%
\BCBL {}\ \BBA {} Zhao%
}{%
Yu%
\ \protect \BOthers {.}}{%
{\protect \APACyear {2021}}%
}]{%
Yuetal2021}
\APACinsertmetastar {%
Yuetal2021}%
\begin{APACrefauthors}%
Yu, Y.%
, He, Y.%
\BCBL {}\ \BBA {} Zhao, X.%
\end{APACrefauthors}%
\unskip\
\newblock
\APACrefYearMonthDay{2021}{}{}.
\newblock
{\BBOQ}\APACrefatitle {Impact of demand information sharing on organic farming
  adoption: An evolutionary game approach} {Impact of demand information
  sharing on organic farming adoption: An evolutionary game approach}.{\BBCQ}
\newblock
\APACjournalVolNumPages{Technological Forecasting and Social
  Change}{172}{}{121001}.
\newblock
\begin{APACrefDOI} \doi{10.1016/j.techfore.2021.121001} \end{APACrefDOI}
\PrintBackRefs{\CurrentBib}

\bibitem [\protect \citeauthoryear {%
Yu%
, He%
, Zhao%
\BCBL {}\ \BBA {} Zhou%
}{%
Yu%
\ \protect \BOthers {.}}{%
{\protect \APACyear {2022}}%
}]{%
yuetal2022}
\APACinsertmetastar {%
yuetal2022}%
\begin{APACrefauthors}%
Yu, Y.%
, He, Y.%
, Zhao, X.%
\BCBL {}\ \BBA {} Zhou, L.%
\end{APACrefauthors}%
\unskip\
\newblock
\APACrefYearMonthDay{2022}{}{}.
\newblock
{\BBOQ}\APACrefatitle {Certify or not? An analysis of organic food supply chain
  with competing suppliers} {Certify or not? an analysis of organic food supply
  chain with competing suppliers}.{\BBCQ}
\newblock
\APACjournalVolNumPages{Annals of Operations Research}{314}{2}{645--675}.
\newblock
\begin{APACrefDOI} \doi{10.1007/s10479-019-03465-y} \end{APACrefDOI}
\PrintBackRefs{\CurrentBib}

\bibitem [\protect \citeauthoryear {%
Zambujal-Oliveira%
}{%
Zambujal-Oliveira%
}{%
{\protect \APACyear {2021}}%
}]{%
zambujal2021}
\APACinsertmetastar {%
zambujal2021}%
\begin{APACrefauthors}%
Zambujal-Oliveira, J.%
\end{APACrefauthors}%
\unskip\
\newblock
\APACrefYearMonthDay{2021}{}{}.
\newblock
{\BBOQ}\APACrefatitle {Supply chain innovation research: A conceptual approach
  of information management with game theory} {Supply chain innovation
  research: A conceptual approach of information management with game
  theory}.{\BBCQ}
\newblock
\APACjournalVolNumPages{Group Decision and Negotiation}{30}{2}{377--394}.
\newblock
\begin{APACrefDOI} \doi{10.1007/s10726-019-09640-7} \end{APACrefDOI}
\PrintBackRefs{\CurrentBib}

\bibitem [\protect \citeauthoryear {%
Zambujal-Oliveira%
\ \BBA {} Fernandes%
}{%
Zambujal-Oliveira%
\ \BBA {} Fernandes%
}{%
{\protect \APACyear {2024}}%
}]{%
zambujal2024}
\APACinsertmetastar {%
zambujal2024}%
\begin{APACrefauthors}%
Zambujal-Oliveira, J.%
\BCBT {}\ \BBA {} Fernandes, C.%
\end{APACrefauthors}%
\unskip\
\newblock
\APACrefYearMonthDay{2024}{}{}.
\newblock
{\BBOQ}\APACrefatitle {The contribution of sustainable packaging to the
  circular food supply chain} {The contribution of sustainable packaging to the
  circular food supply chain}.{\BBCQ}
\newblock
\APACjournalVolNumPages{Packaging Technology and Science}{37}{5}{443--456}.
\newblock
\begin{APACrefDOI} \doi{10.1002/pts.2802} \end{APACrefDOI}
\PrintBackRefs{\CurrentBib}

\bibitem [\protect \citeauthoryear {%
Zanoli%
, Gambelli%
\BCBL {}\ \BBA {} Bruschi%
}{%
Zanoli%
\ \protect \BOthers {.}}{%
{\protect \APACyear {2012}}%
}]{%
Zanolietal2012}
\APACinsertmetastar {%
Zanolietal2012}%
\begin{APACrefauthors}%
Zanoli, R.%
, Gambelli, D.%
\BCBL {}\ \BBA {} Bruschi, V.%
\end{APACrefauthors}%
\unskip\
\newblock
\APACrefYearMonthDay{2012}{}{}.
\newblock
{\BBOQ}\APACrefatitle {Analysis of non-compliances in the organic certification
  system in {Turkey}} {Analysis of non-compliances in the organic certification
  system in {Turkey}}.{\BBCQ}
\newblock
\APACjournalVolNumPages{New Medit}{11}{4}{74--79}.
\PrintBackRefs{\CurrentBib}

\bibitem [\protect \citeauthoryear {%
Zhang%
\ \BBA {} Georgescu%
}{%
Zhang%
\ \BBA {} Georgescu%
}{%
{\protect \APACyear {2022}}%
}]{%
ZhangGeorgescu2022}
\APACinsertmetastar {%
ZhangGeorgescu2022}%
\begin{APACrefauthors}%
Zhang, H.%
\BCBT {}\ \BBA {} Georgescu, P.%
\end{APACrefauthors}%
\unskip\
\newblock
\APACrefYearMonthDay{2022}{}{}.
\newblock
{\BBOQ}\APACrefatitle {Sustainable Organic Farming, Food Safety and Pest
  Management: An Evolutionary Game Analysis} {Sustainable organic farming, food
  safety and pest management: An evolutionary game analysis}.{\BBCQ}
\newblock
\APACjournalVolNumPages{Mathematics}{10}{13}{2269}.
\newblock
\begin{APACrefDOI} \doi{10.3390/math10132269} \end{APACrefDOI}
\PrintBackRefs{\CurrentBib}

\bibitem [\protect \citeauthoryear {%
Zhang%
\ \BBA {} Liu%
}{%
Zhang%
\ \BBA {} Liu%
}{%
{\protect \APACyear {2025}}%
}]{%
Zhangetal2025}
\APACinsertmetastar {%
Zhangetal2025}%
\begin{APACrefauthors}%
Zhang, H.%
\BCBT {}\ \BBA {} Liu, Y.%
\end{APACrefauthors}%
\unskip\
\newblock
\APACrefYearMonthDay{2025}{}{}.
\newblock
{\BBOQ}\APACrefatitle {Impact of quality management system on operational
  efficiency in agro-processing SMEs} {Impact of quality management system on
  operational efficiency in agro-processing smes}.{\BBCQ}
\newblock
\APACjournalVolNumPages{Total Quality Management \& Business
  Excellence}{36}{2}{145--162}.
\newblock
\begin{APACrefDOI} \doi{10.1080/14783363.2025.2502566} \end{APACrefDOI}
\PrintBackRefs{\CurrentBib}

\bibitem [\protect \citeauthoryear {%
J.~Zhao%
, Gerasimova%
, Peng%
\BCBL {}\ \BBA {} Sheng%
}{%
J.~Zhao%
\ \protect \BOthers {.}}{%
{\protect \APACyear {2020}}%
}]{%
Zhaoetal2020}
\APACinsertmetastar {%
Zhaoetal2020}%
\begin{APACrefauthors}%
Zhao, J.%
, Gerasimova, K.%
, Peng, Y.%
\BCBL {}\ \BBA {} Sheng, J.%
\end{APACrefauthors}%
\unskip\
\newblock
\APACrefYearMonthDay{2020}{}{}.
\newblock
{\BBOQ}\APACrefatitle {Information asymmetry, third party certification and the
  integration of organic food value chain in China} {Information asymmetry,
  third party certification and the integration of organic food value chain in
  china}.{\BBCQ}
\newblock
\APACjournalVolNumPages{China Agricultural Economic Review}{12}{1}{20--38}.
\newblock
\begin{APACrefDOI} \doi{10.1108/CAER-05-2018-0111} \end{APACrefDOI}
\PrintBackRefs{\CurrentBib}

\bibitem [\protect \citeauthoryear {%
L.~Zhao%
, Fu%
\BCBL {}\ \BBA {} Bai%
}{%
L.~Zhao%
\ \protect \BOthers {.}}{%
{\protect \APACyear {2025}}%
}]{%
Zhaoetal2025}
\APACinsertmetastar {%
Zhaoetal2025}%
\begin{APACrefauthors}%
Zhao, L.%
, Fu, B.%
\BCBL {}\ \BBA {} Bai, S.%
\end{APACrefauthors}%
\unskip\
\newblock
\APACrefYearMonthDay{2025}{}{}.
\newblock
{\BBOQ}\APACrefatitle {Understanding the Influence of Personalized
  Recommendation on Purchase Intentions from a Self-Determination Perspective:
  Contingent upon Product Categories} {Understanding the influence of
  personalized recommendation on purchase intentions from a self-determination
  perspective: Contingent upon product categories}.{\BBCQ}
\newblock
\APACjournalVolNumPages{Journal of Theoretical and Applied Electronic Commerce
  Research}{20}{1}{32}.
\newblock
\begin{APACrefDOI} \doi{10.3390/jtaer20010032} \end{APACrefDOI}
\PrintBackRefs{\CurrentBib}

\end{thebibliography}
\end{document}